\documentclass{article}
\usepackage{graphicx} 
\usepackage{amsmath}
\usepackage{amssymb}
\usepackage{hyperref}
\usepackage{geometry}
\usepackage{xcolor}
\title{Rips and regular future scenario with Holographic Dark Energy : A comprehensive look}
\author{%
    I. Brevik$^{1}$\thanks{iver.h.brevik@ntnu.no}, Maxim Khlopov$^{2,3,4}$\thanks{khlopov@apc.in2p3.fr}, S.D. Odintsov $^{5,6}$ $\thanks{odintsov@ice.csic.es}$, Alexander V. Timoshkin $^{7,8}$\thanks{alex.timosh@rambler.ru}, Oem Trivedi $^{9}$\thanks{oem.t@ahduni.edu.in} 
}

\date{%
    \small
    $^{1}$Department of Energy and Process Engineering,
    Norwegian University of Science and Technology,
    N-7491 Trondheim, Norway\\
     $^{2}$Research Institute of Physics, Southern Federal University, 344090 Rostov-on-Don, Russia\\
    $^{3}$Virtual Institute of Astroparticle Physics, 75018 Paris, France\\
    $^{4}$Center for Cosmoparticle Physics Cosmion, National Research Nuclear University “MEPHI”, 115409 Moscow, Russia\\
    $^{5}$ Institute of Space Sciences (ICE, CSIC) C. Can Magrans s/n, 08193 Barcelona, Spain\\ 
    $^{6}$ ICREA, Passeig Luis Companys, 23, 08010 Barcelona, Spain\\
    $^{7}$Institute of Scientific Research and Development, Tomsk State Pedagogical University (TSPU),
634061 Tomsk, Russia\\
    $^{8}$ Lab. for Theor. Cosmology, International Centre of Gravity and Cosmos,
Tomsk State University of Control Systems and Radio Electronics (TUSUR), 634050 Tomsk, Russia\\
    $^{9}$International Centre for Space and Cosmology, Ahmedabad University, Ahmedabad 380009, India\\
    \today 
}

\DeclareMathOperator{\sech}{sech}

\usepackage[backend=biber,style=numeric,sorting=none]{biblatex}
\DeclareFieldFormat{eprint}{%
  \texttt{arXiv:\href{http://arxiv.org/abs/#1}{#1}}}

\renewbibmacro*{doi+eprint+url}{
  \printfield{doi}
  \newunit\newblock
  \printfield{eprint}
  \newunit\newblock
  \printfield{url}
}

\renewbibmacro*{in:}{}

\begin{document}

\maketitle

\begin{abstract}
    Interest on the possible future scenarios the universe could have has grew substantially with breakthroughs on late-time acceleration. Holographic dark energy  (HDE) presents a very interesting approach towards addressing late-time acceleration, presenting an intriguing interface of ideas from quantum gravity and cosmology. In this work we present an extensive discussion of possible late-time scenarios, focusing on rips and similar events, in a universe with holographic dark energy. We discuss these events in the realm of the generalized Nojiri-Odintsov cutoff and also for the more primitive holographic cutoffs like Hubble, particle and event horizon cutoffs. We also discuss the validity of the generalized second law of thermodynamics and various energy conditions in these regimes. Our work points towards the idea that it is not possible to have alternatives of the big rip consistently in the simpler HDE cutoffs, and shows the flexibility of the generalized HDE cutoff as well.
\end{abstract}

\section{Introduction}
The discovery of the Universe's late-time acceleration marked a major milestone in cosmology \cite{SupernovaSearchTeam:1998fmf}. Following this breakthrough, substantial work has been dedicated to deciphering this expansion and literature has explored numerous approaches, including conventional methods such as the Cosmological constant \cite{Weinberg:1988cp,Lombriser:2019jia,Padmanabhan:2002ji}, unconventional theories like Modified gravity \cite{Capozziello:2011et,Nojiri:2010wj,Nojiri:2017ncd}, and models where scalar fields drive the late-time cosmic acceleration \cite{Zlatev:1998tr,Faraoni:2000wk,Gasperini:2001pc,Capozziello:2003tk,Capozziello:2002rd,Odintsov:2023weg}. Quantum gravity theories, such as Braneworld cosmology in string theory, loop quantum cosmology, and asymptotically safe cosmology, have also contributed to addressing the cosmic-acceleration puzzle \cite{Sahni:2002dx,Sami:2004xk,Tretyakov:2005en,Chen:2008ca,Fu:2008gh,Bonanno:2001hi,Bonanno:2001xi,Bentivegna:2003rr,Reuter:2005kb,Bonanno:2007wg,Weinberg:2009wa}. Nevertheless, inconsistencies persist, exemplified by the well-known Hubble tension, which underscores the limitations of our current understanding of the Universe and is evidenced by conflicting measurements of the Hubble constant \cite{Planck:2018vyg,riess2019large,riess2021comprehensive}. Therefore, the current epoch of the Universe raises profound questions, with the potential for significant advancements in gravitational physics to enhance our understanding of cosmology.
\\
\\
Among the various proposed frameworks to address the challenges of late-time cosmic acceleration, the holographic principle \cite{tHooft:1993dmi,Susskind:1994vu} has emerged as a compelling concept in cosmology. This principle asserts that a system’s entropy is governed not by its volume but rather by its surface area \cite{Bousso:1999xy}. Recently, this approach to dark energy (DE) has garnered significant attention, particularly in light of new results from DESI \cite{DESI:2024lzq,DESI:2024mwx,DESI:2024uvr}, which suggest that deviations from the standard $\Lambda$CDM model may remain a viable possibility. Early investigations into holographic dark energy (HDE) by \cite{Cohen:1998zx} explored this idea using a quantum field theory perspective, demonstrating that a connection between short-distance and long-distance cutoffs arises from constraints imposed by black hole formation. Specifically, if $\rho$ represents the quantum zero-point energy density corresponding to a short-distance cutoff, the total energy within a region of size \( L \) must not exceed the mass of a black hole of the same size. This yields the constraint \( L^3 \rho \leq L M_{\text{pl}}^2 \). The largest permissible value of the infrared cutoff, denoted as \( L_{_{\text{IR}}} \), saturates this inequality, leading to the relationship

\begin{equation}\label{simphde}
\rho = 3 c^2 L_{_{\text{IR}}}^{-2},
\end{equation}

where \( c \) is a free parameter, and we adopt units with \( m_{p} = 1 \footnote{We emphasize that we will consistently use the \( m_{p}=1 \) units throughout this paper, including for other holographic dark energy models.} \).

This principle has found wide-ranging applications in cosmology, particularly in explaining the late-time acceleration era attributed to dark energy, under the framework of holographic dark energy (HDE). For an in-depth review, see \cite{Wang:2016och}. In this framework, the infrared cutoff \( L_{_{\text{IR}}} \) plays a crucial role, representing the IR cutoff for a particular HDE model. Over the past few years, holographic dark energy has been investigated extensively from a variety of perspectives \cite{Nojiri:2017opc,Oliveros:2022biu,Granda:2008dk,Khurshudyan:2016gmb,Wang:2016och,Khurshudyan:2016uql,Belkacemi:2011zk,Zhang:2011zze,Setare:2010zy,Nozari:2009zk,Sheykhi:2009dz,Xu:2009xi,Wei:2009au,Setare:2008hm,Saridakis:2007wx,Setare:2006yj,Felegary:2016znh,Dheepika:2021fqv,Nojiri:2005pu,Nojiri:2021iko,Nojiri:2020wmh,Trivedi:2024dju,Trivedi:2024rhp}.

In addition, several alternative formulations of HDE have been proposed over the years. For instance, Tsallis holographic dark energy (THDE) incorporates Tsallis entropy corrections to the conventional Boltzmann-Gibbs entropy, yielding the energy density

\begin{equation}\label{rtsa}
\rho_{\Lambda} = 3 c^2 L^{-(4 - 2\sigma)},
\end{equation}

where \( \sigma \) is the Tsallis parameter, assumed to be positive \cite{Tavayef:2018xwx}, and the standard HDE is recovered in the limit \( \sigma \to 1 \). Tsallis entropy corrections are grounded in quantum gravity, providing novel insights into cosmological dynamics.

Similarly, Barrow's reformulation of the Bekenstein-Hawking entropy inspired the Barrow holographic dark energy (BHDE) model. This model modifies the HDE energy density to take the form

\begin{equation}\label{rbar}
\rho_{\Lambda} = 3 c^2 L^{\Delta - 2},
\end{equation}

where \( \Delta \) is a deformation parameter capped at \( \Delta = 1 \). In the limit \( \Delta \to 0 \), the standard HDE is recovered \cite{Saridakis:2020zol}. Barrow’s insights, inspired by fractal-like features seen in black hole structures (partially influenced by Covid-19 virus imagery), demonstrated that quantum-gravitational effects might induce fractal properties in black hole geometries. This leads to entropy expressions with finite or infinite area and corresponding modifications to the energy density, as reflected in the above formulation. In recent times, a substantial body of literature has emerged focusing on the exploration of various types of singularities that may arise in the future evolution of the Universe. The detection of late-time acceleration has significantly propelled such investigations \cite{Nojiri:2004ip,Nojiri:2005sr,Nojiri:2005sx,Bamba:2008ut,trivedi2022finite,trivedi2022type,odintsov2015singular,odintsov2016singular,oikonomou2015singular,nojiri2015singular,odintsov2022did,Trivedi:2023aes,Trivedi:2023rln,Trivedi:2023wgg,Trivedi:2023zlf,deHaro:2023lbq}. A particularly interesting class of such future events are rip scenarios, where the universe may proceed towards progressive disintegration in various capacities.  An interesting question to ponder then is to ask which rip scenarios would be feasible in the vast array of holographic dark energy models. A recent work \cite{Trivedi:2024dju} addressed this in the context of HDE models with the Granda-Oliveros cutoff \cite{Granda:2008dk}, finding that the overarching possibility is that of the big rip while its alternatives are harder to come by. In this work we would like to go even further and take into account all other possible forms of IR cutoff for HDE theories and consider the feasibility of various rip and rip-like scenarios here. We would discuss these events with the primitive HDE cutoffs of the Hubble horizon, particle horizon and event horizon types, considering these events with these cutoffs in the conventional HDE model \eqref{simphde}, Tsallis model \eqref{rtsa} and Barrow model \eqref{rbar}. We would also discuss about the generality of the Nojiri-Odintsov cutoff (N-O cutoff), and how it can facilitate multiple scenarios in this case as well. In section II we shall give a brief overview of rip scenarios and some of the methods which have been discussed in a bid to avoid them. In section III we shall discuss the generalized Holographic dark energy paradigm with the N-O cutoff and look at how various future scenarios can be accommodated in it. In section IV we shall discuss about such future events with the Hubble horizon cutoff, while in section V and VI we shall discuss it for the particle and event horizon cutoffs. In section VII we shall discuss about the thermodynamic aspects and the validity of energy conditions in various scenarios, while we conclude our work in section VIII. 
\\
\\
\section{Rips, cosmological singularities and their avoidance procedures}
An interesting variety of rip scenarios have been put forward in recent years. Even besides rip models, a rich literature on future cosmological singularities has also developed and we can summarize the rip possibilities and other singularities as follows:
\begin{itemize}
\item Big Rip (Type I singularity): A well-known scenario where, as $t \to t_{f}$ (with $t_{f}$ finite), both the effective energy density and pressure density of the universe diverge, $\rho_{eff} \to \infty$, $p_{eff} \to -\infty$, and the Hubble parameter also diverges, $H \to \infty$ \cite{Caldwell:2003vq}. This results in a scenario of universal destruction, where everything in the universe undergoes progressive disintegration \cite{Caldwell:2003vq}.
\item Little Rip (LR): In this scenario, the density, pressure, and Hubble parameter become infinite as $t \to \infty$ \cite{Frampton:2011sp}. In this case, all bound structures are eventually torn apart, but there is no finite-time singularity.
\item Pseudo Rip (PR): Here, $H$ increases monotonically as $t \to \infty$, but it is bounded from above by the value $H_{\infty}$ so that $H \to H_{\infty}$ as $t \to \infty$ \cite{Frampton:2011aa}.
\item Sudden singularity (Type II singularity) : In the case of type II singularities, the pressure density diverges and so does the derivatives of the scalar factor from the second derivative onwards \cite{Barrow:2004xh}. The weak and strong energy conditions hold for this singularity. Also known as quiescent singularities\footnote{This name originally appeared in contexts related to non-oscillatory singularities \cite{Andersson:2000cv}.}, a special case of this is the big brake singularity \cite{Gorini:2003wa}.
\item Big Freeze (Type III singularity): In this case, the derivative of the scale factor from the first derivative onward diverges. These were detected in generalized Chaplygin gas models \cite{bouhmadi2008worse}. 
\item Generalized sudden singularities (Type IV singularities) : These are finite time singularities with finite density and pressure instead of diverging pressure. In this case, the derivative of the scale factor diverges from a derivative higher than the second \cite{Bamba:2008ut,Nojiri:2004pf}.  
\end{itemize}
Note that the Little Rip and Pseudo Rip are infinite future singularities while the others are finite-time singularities.These singularities have been found and discussed in a huge array of cosmological models, ranging from simple $\Lambda$CDM models, to exotic models of all kinds like Chaplygin gas, scalar field dark energy models, modified gravity theories, Braneworld cosmologies etc. \cite{deHaro:2023lbq,Trivedi:2023zlf}. With such singularities being found in various models, a natural interest also grew in finding ways to avoid, delay or outright remove these singularities. An approach to mitigating or delaying singularities involves considering the effects of conformal anomalies. The impact of quantum backreaction of conformal matter on Type I, Type II, and Type III singularities was explored in \cite{Nojiri:2005sx,Nojiri:2004ip,Nojiri:2000kz}. Around singularity time \( t = t_s \), the curvature of the universe becomes large, although the scale factor \( a \) remains finite for Type II and III singularities. Quantum corrections, which typically involve powers of curvature or higher derivative terms, become significant near the singularity.
\\
\\
Conformal anomalies occur due to the presence of one-loop vacuum contributions from various matter fields, breaking the conformal invariance of the matter action in the renormalized action. The trace of the energy-momentum tensor in a conformally invariant theory is zero classically, but renormalization leads to a non-zero trace, known as the conformal anomaly. The conformal anomaly can be expressed as \cite{Nojiri:2005sx}:

\begin{equation}
T_{A} = b \left( F + \frac{2}{3} \Box R \right) + b^{\prime} G + b^{\prime \prime} \Box R
\end{equation}

where \( T_{A} \) is the trace of the stress-energy tensor, \( F \) is the square of the 4D Weyl tensor, and \( G \) is the Gauss-Bonnet curvature invariant, defined as:

\begin{equation}
F = \frac{1}{3} R^2 - 2 R_{ij} R^{ij} + R_{ijkl} R^{ijkl}
\end{equation}

\begin{equation}
G = R^2 - 4 R_{ij} R^{ij} + R_{ijkl} R^{ijkl}
\end{equation}

The coefficients \( b \) and \( b^{\prime} \) are given by:

\begin{equation}
b = \frac{N + 6 N_{1/2} + 12 N_{1} + 611 N_{2} - 8 N_{HD}}{120 (4 \pi)^2}
\end{equation}

\begin{equation}
b^{\prime} = - \frac{N + 11 N_{1/2} + 62 N_{1} + 1411 N_{2} - 28 N_{HD}}{360 (4 \pi)^2}
\end{equation}

where \( N \) is the number of scalar fields, \( N_{1/2} \) is the number of spinor fields, \( N_{1} \) is the number of vector fields, \( N_{2} \) (which can be 0 or 1) represents gravitons, and \( N_{HD} \) denotes higher derivative conformal scalars. Typically, \( b > 0 \) and \( b^{\prime} < 0 \) for ordinary matter, except for higher derivative conformal scalars, while \( b^{\prime \prime} \) can be arbitrary.

Quantum effects due to the conformal anomaly act as a fluid with energy density \( \rho_{A} \) and pressure \( p_{A} \). The total energy density is \( \rho_{tot} = \rho + \rho_{A} \). The conformal anomaly (trace anomaly) can be given by:

\begin{equation}
T_{A} = - \rho_{A} + 3 p_{A}
\end{equation}

Using the continuity equation, we have \cite{Nojiri:2005sx}:

\begin{equation} \label{4.8}
T_{A} = -4 \rho_{A} - \frac{\dot{\rho_{A}}}{H}
\end{equation}

In standard cosmology, \( \rho_{A} \) can be expressed as an integral in terms of \( T_{A} \):

\begin{equation}
\rho_{A} = - \frac{1}{a^4} \int a^4 H T_{A} dt
\end{equation}

Furthermore, \( T_{A} \) can be written in terms of the Hubble parameter:

\begin{equation}
T_{A} = -12 b \dot{H}^2 + 24 b^{\prime} (-\dot{H}^2 + H^2 \dot{H} + H^4) - (4b + 6 b^{\prime \prime}) (H^{(3)} + 7 H \ddot{H} + 4 \dot{H}^2 + 12 H^2 \dot{H})
\end{equation}

Using this, \( \rho_{A} \) can be derived to account for conformal anomaly effects near singularities. One can then find out that this setup can mitigate quite a few future rips and singularities. Consider the EOS being taken of the form  $p=-\rho - f(\rho)$ with $f\rho) = B \rho^{\beta}$, where B and $\beta$ are positive constants. This is a form which has been shown to be a valid cosmology to give quite a few future singularities including the big rip, and these conformal anomaly effects are enough to alleviate even the big rip in this case \cite{Nojiri:2004ip}. The conformal anomaly effects do not always work \cite{Trivedi:2022svr}, however, they are nonetheless quite effective. There are other approaches to mitigating these singularities as well, one such being about considering in a modified gravity background \cite{Nojiri:2017ncd,Nojiri:2010wj,Nojiri:2008fk}. One can even consider cosmologies with varying physical constants, like speed of light or the gravitational constant\cite{barrow1999cosmologies}. It has been shown to regularize cosmological singularities in certain scenarios \cite{dkabrowski2016new,Dabrowski:2012eb,Leszczynska:2014xba,Salzano:2016pny}. A common theme in these approaches is that they consider ways of modifying the background gravitational theory or other fundamental essentials of standard cosmology in order to alleviate the issues of singularities. We would now like to discuss in the next section how holographic dark energy approaches this scenario, and how it can alleviate rips and other singularities in the generalized HDE scenario. 
\\
\\
\section{Generalized holographic dark energy and late-time scenarios}
There has been a flurry of work on the paradigm of Holographic dark energy theories in recent times. This has resulted in a diversified literature where various forms of the energy density of the HDE alongside various potential cutoff scales have been discussed. The initial HDE proposals had dealt with the simple IR cutoff where the scale L was identified with \begin{equation}\label{hcut}
    L \to H^{-1}
\end{equation} 
where H is the Hubble parameter, with this cutoff being regarded as the Hubble horizon cutoff. More involved cutoffs included those where the cutoff was identified with the particle Horizon \begin{equation} \label{parcut}
    L_{p} = a \int_{0}^{t} \frac{dt}{a}
\end{equation}
or with the future cosmological event horizon \begin{equation} \label{eventcut}
    L_{f} = a \int_{a}^{\infty} \frac{dt}{a}
\end{equation}
These different cutoff choices, while popular, had their own set of issues as well. In the case of the Hubble horizon for example, it was found that this particular scale resulted in an equation of state approaching zero, failing to contribute significantly to the current accelerated expansion of the universe. In the particle horizon approach the equation of state parameter higher than $-\frac{1}{3}$ but the struggle of explaining the present acceleration remained as it was.
In the event horizon case one was getting the desired results, although at the cost of issues with regards to  causality, posing a significant obstacle to its viability \footnote{Note that these observations are in regards to these cutoffs being used in the conventional HDE energy density setup with $\rho_{DE} \to L^{-2}$ \eqref{simphde}, however these worries have been expressed in other HDE models with these cutoffs as well.}. Another choice which was then proposed for the HDE cutoff was the Granda-Oliveros cutoff, where one has \begin{equation} \label{gocut}
    L = \left( \alpha H^2 + \beta \dot{H} \right)^{\frac{-1}{2}}
\end{equation} 
This cutoff \footnote{Note that these models have also been referred to as "generalized ghost dark energy" as well \cite{cai2011notes}} was proposed in order alleviate the issues that the former cutoffs and while it does give desired results in a large number of cases, arguably better performing than the previous cutoffs, one can still have other issues like a prominent one with regards to the classical stability of the HDE models. This comes around due to the fact that various HDE models, in all 4 cutoffs as discussed above, have been seen to have negative squared sound speeds which is indeed a worrisome feature. There are other issues where the Granda-Olvieros can, in principle, perform better than the other cutoffs as well like with regards to validity of energy conditions and the validity of the generalized second law of thermodynamics in models with these cutoffs. 
\\
\\
A general observation one can make here is that the results in HDE scenarios become better suited given a more suitable and dynamic interplay of cosmological parameters in the cutoff scale. Another key point to note is that a prominent  issue of the original holographic dark energy model \cite{li2004model}, where the infrared cutoff was chosen as the size of the event horizon, was that the corresponding FLRW equations often do not correspond to any covariant gravity theory and even may not predict universal acceleration appropriately. With these notions in mind, one turns their attention to the generalized holographic dark energy scenario, where L is identified by the N-O given as \cite{Nojiri:2005pu} \begin{equation} \label{gencut}
    L = L(L_{p}, \dot{L_{p}},\ddot{L_{p}},...,L_{f},\dot{L_{f}},\ddot{L_{f}},...,H,\dot{H},\ddot{H},..)
\end{equation}
As is clear from \eqref{gencut}, this cutoff scale encapsulates all the previous proposals we have discussed. All of the Hubble, particle, event and Granda-Oliveros cutoff scales turn into special cases of this generalized N-O cutoff scheme. Given the general and quite unique setup of the cutoff, it is hence of interest to study late universe events in this paradigm as well. Recently in \cite{Trivedi:2024dju} the authors showed discussed about the possibility of various rip events in quite a few HDE models with the Granda-Oliveros cutoff. The authors were able to show that the overarching possibility in such models is the Big rip, which can happen in almost any HDE model while a pseudo rip is also a real possibility in some cases. The curious feature is that the little rip occurs is virtually impossible to have in HDE scenarios with this cutoff. Given the fact that the G-O cutoff in general presents a more complete HDE scenario than the simple Hubble, particle and event horizon cases, it is indeed an interesting result. Building on this, however, one can then also ask about the possibility of various rip events in the the N-O cutoff and we would like to briefly discuss this as well. 
\\
\\
For the moment, we shall consider the conventional HDE energy density \eqref{simphde} i.e. we are not considering any extended HDE scenarios like Tsallis, Barrow etc. Considering that in the very late universe, dark energy is the dominant material component one can say, without loss of generality, that $\rho_{DE} > \rho_{m}, \rho_{r}$, where $\rho_{m}$ and $\rho_{r}$ mean the matter and radiation energy densities, respectively. With this in mind, one can then write using the Friedmann equation \begin{equation}
    H^2 = \frac{\rho}{3} \sim \frac{\rho_{DE}}{3}
\end{equation} while being given the cutoff \eqref{gocut} \begin{equation} \label{simpintegral}
   \int_{H_{i}}^{H_{f}} \frac{ \beta c^2  dH}{H^2 ( 1- \alpha c^2)} = \frac{\beta  c^2 \left(\frac{1}{H_{i}}-\frac{1}{H_{f}}\right)}{1-\alpha  c^2} = \int_{t_{i}}^{t_{f}} dt.
\end{equation}
From this it is pretty clear that in the context of this HDE model with the Granda-Oliveros cutoff, only a big rip scenario can be allowed for. In general now, one can find an integral of the form \cite{Trivedi:2024dju} \begin{equation} \label{mainintegral}
    \int_{H_{i}}^{H_{f}} \frac{d H}{g(H)} = \int_{t_{i}}^{t_{f}} dt
\end{equation}
The above integral can come from various HDE models in the G-O cutoff paradigm, including various modifications of the simple HDE energy density. For $H_{f} \to \infty$, if the integral on the LHS \eqref{mainintegral} then we have $t_{f} \to \infty$ as well, which means that a little rip could be a possibility. Similarly if the LHS stays finite for $H_{f} \to \infty$ then a big rip could happen while if the integral diverges for a finite $H_{f}$ then we have a pseudo rip scenario. However, the above integral can also be obtained from the generalized N-O cutoff as well and one can work through the corresponding integrals with the same conditions to talk of rips in those cases as well. Here we would like to make the case that in the generalized cutoff scale events which are "forbidden" in the G-O cutoff scale can also take place here. Consider for example a more involved form of \eqref{gocut}, given as \begin{equation} \label{cut1}
    L = (\alpha_{1} H + \alpha_{2} H^2 + \alpha_{3} H^3 + \beta \dot{H})^{-\frac{1}{2}}
\end{equation} where $\alpha_{1}, \alpha_{2}$ and $\alpha_{3}$ are constants. Note that \eqref{cut1} becomes the Granda-Oliveros cutoff in the limit $\alpha_{1}, \alpha_{3} \to 0$. In this case, we have the corresponding integral on the LHS as in \eqref{mainintegral} as \begin{equation} \label{cut1int}
   \int_{H_{i}}^{H_{f}} \frac{dH}{-\alpha_{1} H+ (1-\alpha_{2}) H^2-\alpha_{3} H^3} \, = \frac{\beta}{2 \alpha_{1}} \Bigg[ \frac{2 (\alpha_{2}-1) \tan ^{-1}\left(\frac{\alpha_{2}+2 \alpha_{3} H-1}{\sqrt{4 \alpha_{1} \alpha_{3}-\alpha_{2}^2+2 \alpha_{2}-1}}\right)}{\sqrt{4 \alpha_{1} \alpha_{3}-\alpha_{2}^2+2 \alpha_{2}-1}}+\ln \left( \frac{(\alpha_{1}+H (\alpha_{2}+\alpha_{3} H-1))}{H^2} \right) \Bigg] + \text{constant} 
\end{equation} We see that for $$ H \to \frac{\sqrt{4 \alpha_{1} \alpha_{3}-\alpha_{2}^2+2 \alpha_{2}-1}-\alpha_{2}+1}{2 \alpha_{3}} $$ the integral in \eqref{cut1int} diverges, which means this is the H takes as $t_{f} \to \infty$. This means that in a dark energy model with monotonically increasing DE energy density, H takes a finite value at infinite times which signifies that we could have a pseudo rip on our hands here. This goes to show that even form a seemingly extended form of the G-O cutoff as in \eqref{cut1}, which is again just yet another formulation of the N-O, can give us a pseudo rip scenario even when the form of the energy density is just of the conventional form. We may not even need to just rely on the integral approach in certain cases and can see more exact formulations as well. Consider the following form of \eqref{gencut} for example \cite{Nojiri:2017opc} \begin{equation} \label{cut2}
    \frac{c}{L} = \frac{1}{L} = \left( 1 - \frac{1}{t_{0} h_{0}} \right) \frac{1}{L_{f}} + \frac{2 h_{0}}{t_{0} h_{1}} + \frac{h_{1}}{t_{0}} \left( 1 - \frac{h_{0}^2}{h_{1}^2} \right) L_{f}
\end{equation}
where we have henceforth set $c=1$ (which is valid in these generalized schemes), with $h_{0}, t_{0}, h_{1}$ being constants with $h_{0} > h_{1} $. In this case, one can use the Friedmann equation to get the Hubble parameter to be in the form  \begin{equation} \label{ht1}
    H(t)= \frac{h_{1} }{t_{0} \left(h_{0}-h_{1} \tanh \left(\frac{t}{t_{0}}\right)\right) \cosh^2{\left(\frac{t}{t_{0}}\right)} }+h_{0}-h_{1} \tanh \left(\frac{t}{t_{0}}\right)
\end{equation} 
One sees in this case as well that as $t \to \infty$, $H \to h_{0} - h_{1}$, which is again a constant value and one sees again a situation of a pseudo rip. In this case one can also write \begin{equation} \label{dotr1}
    \dot{\rho} = \frac{2 \rho \dot{L_{f}}}{H \sqrt{\Omega_{d}}} \left( \alpha_{2} - \frac{\alpha_{0}}{L_{f}^2} \right)
\end{equation}
where $\Omega_{d}$ is the fractional energy density parameter for DE. One can also find that \begin{equation}
    \frac{1}{L_{f}} = h_{0}-h_{1} \tanh \left(\frac{t}{t_{0}}\right)
\end{equation}
One can use this form of $L_{f}$ in \eqref{dotr1} and find that even other forms of cosmological singularities. Besides the Big Rip(Type I) and Big Freeze (Type III) singularities, there are other interesting future singularities that one also sees in various cosmological models, which include sudden (type II) and generalized sudden singularities (type IV), as we had discussed before. One can write using \eqref{ht1} that \begin{equation}
    \dot{H} = \frac{1}{t_{0}^2} \left( \frac{(h_{1}-h_{0}) (h_{0}+h_{1})}{\left(h_{0} \cosh \left(\frac{t}{t_{0}}\right)-h_{1} \sinh \left(\frac{t}{t_{0}}\right)\right)^2}+(1-h_{1} t_{0}) \sech^2\left(\frac{t}{t_{0}}\right) \right)
\end{equation}
\begin{equation}
    \ddot{H} = \frac{1}{t_{0}^3} 2 \left(\frac{(h_{0}-h_{1}) (h_{0}+h_{1}) \left(h_{0} \sinh \left(\frac{t}{t_{0}}\right)-h_{1} \cosh \left(\frac{t}{t_{0}}\right)\right)}{\left(h_{0} \cosh \left(\frac{t}{t_{0}}\right)-h_{1} \sinh \left(\frac{t}{t_{0}}\right)\right)^3}+(h_{1} t_{0}-1) \tanh \left(\frac{t}{t_{0}}\right) \sech^2\left(\frac{t}{t_{0}}\right)\right)
\end{equation}
\\
\\
We note interestingly that for no values of t, either finite or infinite, does the Hubble parameter or its derivatives diverge in the above cases. Furthermore, the expressions for the pressure and the energy density, which we do not reproduce here owing to their length, show the same behaviour. This shows that even type II and type IV singularities do not occur for this cutoff choice. Another example of a N-O cutoff is \begin{equation} \label{cut3}
    L = (\beta \dot{H}^2)^{-\frac{1}{2}}
\end{equation} 
Using this form of the cutoff, we end up with the Hubble parameter as \begin{equation}
    H(t) = H_{0} e^{\frac{t-t_{0}}{\sqrt{\beta}}}
\end{equation}
This form of the cutoff can accommodate a Little rip scenario as H diverges here at infinite times. The reason we have considered these cases here is in order to have a gauge on whether or not the more generalized HDE cutoff can provide us with more flexibility in allowing for more late-time events when compared with the G-O cutoff. It indeed seems to be the case here, as we have seen with the cutoffs in \eqref{cut1},\eqref{cut2} and \eqref{cut3}, and how they could naturally allow for pseudo or little rip scenarios even with the conventional HDE energy density \eqref{simphde}. The rip possibilities one can ponder with the N-O cutoffs in extended HDE models like Tsallis, Barrow, Kaniadakis etc. would then be even more interesting to study. It is so as one can even be looking at the possibility that, given certain parameter ranges for the free parameters of these regimes, all three of these rip scenarios could be allowed for or avoided as well in the N-O scheme. This reinforces the point which was briefly mentioned in  \cite{Trivedi:2024dju}, which was about the possibility of the N-O cutoff in general representing the most viable way to have various late time scenarios. We can make a comment in general, based on the discussions of the cutoffs above. Consider depicting two kinds of generalized N-O cutoffs, one being the "Hubble type" \begin{equation} \label{hubtype}
    L = L(H, \dot{H}, \ddot{H}....)
\end{equation}
The other being the "P-E type" \begin{equation} \label{petype}
    L= L(L_{p},\dot{L_{p}},\ddot{L_{p}}...,L_{f},\dot{L_{f}},\ddot{L_{f}}....)
\end{equation}
One would realize that the Hubble type \eqref{hubtype} encapsulate the Hubble horizon \eqref{hcut} and the G-O cutoff \eqref{gocut} as special cases while the P-E type \footnote{By "P-E" we aim to acronymize the Particle and Event horizon type into a single phrase, as this cutoff represents all possible combinations of these scales.} encapsulates all possible cases with the particle horizon \eqref{parcut} and event horizon \eqref{eventcut} cutoffs. Of course, both these formulations are just ways to rephrase the N-O cutoff \eqref{gencut} but it is important to make the distinction here with regards to how rips and singularities come by. In general we would like to comment that one would observe models with cutoffs of the Hubble type \eqref{hubtype}, like \eqref{cut1}-\eqref{cut3}, more easily produce scenarios with rips, singularities and similar futures, given the more direct correspondence of $\rho \to f(H)$ in that case. Such a correspondence leads to cases where any blow ups in H would lead to blow ups in the densities as well, either in finite or infinite time and it would be more difficult to avoid rips and singularities in these cases. Furthermore, as mentioned before, it could very well be the case that if one considers more extended forms of the energy density \eqref{simphde} then the free parameters associated with the models could define ranges in their parameter space where rips and similar events could both be accommodated and avoided, but strictly speaking from the conventional HDE point of view the Hubble type cutoffs would be more difficult in terms of avoiding such blowups. On the other hand, the cutoffs of the P-E type \eqref{petype} like \eqref{petype} would lead to more non-singular and rip free universes. Indeed it has well known that HDE models with simple particle or event horizon cutoffs readily allow for big rip scenarios, but our comment concerns cases where one considers general functions of these cutoffs as in \eqref{petype}. One can say that eventually the correspondence between $\rho \to f(H)$ would stay in both cutoff bifurcations, but differences come about more clearly by how the DE energy density $\dot{\rho}$ evolves in cases of the Hubble types and of the P-E type. So in general one may observe that cosmologies with N-O cutoff of the form \eqref{petype} are more non-singular than those of the form \eqref{hubtype}.    
\\
\\
But going a step back, one is left to wonder whether if the GO cutoff is not able to support various ends of the universe, could the more simple cutoffs do? One can indeed argue that in general we would not want to have rips etc. in the universe, but it has already been discussed frequently in the literature that big rips can occur in various HDE models so it is natural to ask whether alternatives of the big rip and alike events can also be consistently plausible too. Another way to study these late-time events would be by utilizing the ansatz' for these scenarios, and using them to find the forms of relevant observational and theoretical parameters to ascertain the consistency of these events in various models. We would like to now employ this approach to study about the status quo of various late time events with the Hubble, Particle and Event horizon cutoff schemes. In order to make the analysis more extensive, we shall also consider the Tsallis and Barrow models alongside the conventional HDE model in our work, which in total provides us with 9 models ( prospectively) to look at. In the subsequent sections we would be discussing about various subtleties of rips and similar events in such models and will try to gauge the flexibility of the simpler cutoffs in addressing various late-time events in the universe.
\\
\\
\section{Future events with Hubble horizon cutoff}
In all the ensuing work, we shall consider an interacting dark sector. One can explore an interaction between dark energy and dark matter by incorporating the following continuity equations:
\begin{equation} \label{contd}
\dot{\rho}_{\Lambda} + 3 H \rho_{\Lambda} (1+ w_{\Lambda}) = - Q 
\end{equation}

\begin{equation} \label{contm}
\dot{\rho}_{m} + 3 H \rho_{m} (1+ w_{m}) = Q
\end{equation}

In this context, $\rho_{\Lambda}$ and $\rho_m$ denote the energy densities of dark energy and dark matter, respectively. The symbol $w_{\Lambda}$ stands for the equation of state parameter for dark energy, while $w_m$ represents the equation of state parameter for dark matter. The interaction between these two components is described by the term $Q$. It is important to specify here that the sign infront of Q in RHS in the above equations specifies the kind of interaction in the dark sector, with -Q on the RHS in \eqref{contd} and +Q on the RHS of \eqref{contm} showing that we consider a scenario where gradually $\rho_{m}$ is decaying into $\rho_{d}$. If the signs were flipped in the continuity equations above, then the reverse decay from DE to DM would instead be seen.  

To simplify our analysis, we focus exclusively on the contributions of dark energy and dark matter to cosmology. The interaction term $Q$ can be expressed in either linear or nonlinear forms. Examples include:

\begin{equation} \label{q1}
Q = 3Hb (\rho_{\Lambda} + \rho_{m})
\end{equation}

\begin{equation} \label{[q2]}
Q = 3Hb \frac{\rho_{\Lambda}^2}{\rho_{\Lambda} \rho_{m}}
\end{equation}

In these equations, $b$ is a coefficient that determines the strength of the interaction. The different linear and nonlinear forms of $Q$ allow for an exploration of various dynamics within the interacting dark sectors. Before considering any specific ansatz, it is crucial to perform some general analysis that is independent of any particular ansatz. Additionally, we define the fractional energy densities for dark energy (DE) and dark matter (DM) as follows:

\begin{equation} \label{omegad} 
\Omega_{\Lambda} = \frac{\rho_{\Lambda}}{3 H^2}
\end{equation}

\begin{equation} \label{omegam}
\Omega_{m} = \frac{\rho_{m}}{3 H^2}
\end{equation}

Here, $\rho_{\Lambda}$ is dependent on the specific model of holographic dark energy (HDE) chosen. We shall employ various ansatz' here for the Hubble parameter to study the little rip and pseudo rip scenarios (see \cite{Frampton:2011rh}) with a little rip ansatz being\begin{equation} \label{lr1}
    H(t)= H_{0} \exp(\lambda t ) 
\end{equation}
While another variant for the little rip ansatz is \begin{equation} \label{lr2}
    H(t) = H_{0} \exp (C \exp(\lambda t))
\end{equation}
The ansatz for the Pseudo rip can be given as \begin{equation} \label{pr}
    H(t) = H_{0} - H_{1} \exp(-\lambda t)
\end{equation} 
where $C,H_{0},H_{1}$ and $\lambda$ are positive constants with $H_{0} > H_{1}$. While these ansatz' are generally applicable, in \cite{Astashenok:2012iy} it was studied how the ansatz should be developed for the particular case of Braneworld cosmologies. In such a scenario we have, the ansatz for the case of a Little rip, asymptotic dS (Pseudo rip) and Big Freeze in the scenario of a Braneworld background are given as, 
\begin{equation} \label{blr}
    a(t)= a_{0} \exp \left[ \frac{\lambda_{b}}{3 \alpha^2} \cosh \left( \sqrt{\frac{3 \alpha^3}{2 \lambda_{b}}} t \right)  \right]
\end{equation}
\begin{equation} \label{ads}
    a(t) = a_{0} \exp \left[ -\frac{\beta^2}{2} \left[ \frac{1}{\cos \eta_{0}} - \sqrt{1 + \left( \tan \eta_{0} + \frac{\sqrt{\alpha}}{\beta} t \right)^2} \right] \right]
\end{equation}
\begin{equation} \label{bf}
    a(t)=a_{f} \exp\left[ - \frac{\lambda_{b}}{3 \alpha^2} \left(1 - \frac{3 \alpha^3 t^2}{2 \lambda_{b}} \right)^{1/2} \right]  
\end{equation}
where $a_{0}, a_{f},\alpha$ are positive constants while $\lambda_{b}$ is the brane tension. In the case of the Asymptotic dS \eqref{ads} $\beta$ and $\eta_{0}$ are also positive constants, with $\beta$ being in terms of $\lambda_{b}$ and $\alpha$. Also it is worth noting that the Little rip \eqref{blr} and Big Freeze \eqref{bf} cases are possible with positive brane tension $\lambda_{b} > 0$ while the asymptotic dS case occurs with a negative brane tension $\lambda_{b} < 0$. 
\\
\\
We start off with the Hubble horizon cutoff where we have \begin{equation*}
    L \to H^{-1}
\end{equation*}
To start our analysis consider the usual HDE model, \eqref{simphde} from where we can write \begin{equation}
    \dot{\rho} = -2 \rho \frac{\dot{L}}{L} 
\end{equation}
In the case of the cutoff \eqref{hcut},  we have \begin{equation}
    - \frac{\dot{L}}{L} = \frac{\dot{H}}{H}
\end{equation}
From which we end up with the expression \begin{equation} \label{hrhonormal}
    \dot{\rho} = \frac{2 \dot{H} \rho}{H}
\end{equation}
We can then use it in the continuity equation \eqref{contd} to get the equation of state parameter for dark energy as \begin{equation} \label{wrhonormal}
    w = - 1 - \frac{1}{3 H \Omega_{d}} \left( \frac{Q}{3 H^2} + \frac{2 \dot{H} \Omega_{d}}{H} \right)
\end{equation}
Similarly for the case of the Tsallis paradigm with \eqref{rtsa}, we end up with \begin{equation} \label{hrhotsallis}
    \dot{\rho} = \frac{(4-2\sigma) \dot{H} \rho}{H}
\end{equation}
\begin{equation} \label{wrhotsallis}
    w = - 1 - \frac{1}{3 H \Omega_{d}} \left( \frac{Q}{3 H^2} + \frac{(4-2\sigma) \dot{H} \Omega_{d}}{H} \right)
\end{equation}
While in the Barrow case with \eqref{rbar}
We have \begin{equation} \label{hrhobarrow}
    \dot{\rho} = \frac{(2- \Delta) \dot{H} \rho}{H}
\end{equation}
\begin{equation} \label{wrhobarrow}
    w = - 1 - \frac{1}{3 H \Omega_{d}} \left( \frac{Q}{3 H^2} + \frac{(2- \Delta) \dot{H} \Omega_{d}}{H} \right)
\end{equation}
We can now perform some plots in this regard. Firstly we consider the ansatz \eqref{lr1} and we plot the EOS parameter and squared sound speed, given by \begin{equation}
    v_{s}^2 = w + \frac{\dot{w} \rho}{\dot{\rho}}
\end{equation} 
\begin{figure}[!ht]
    \centering
    \includegraphics[width=1\linewidth]{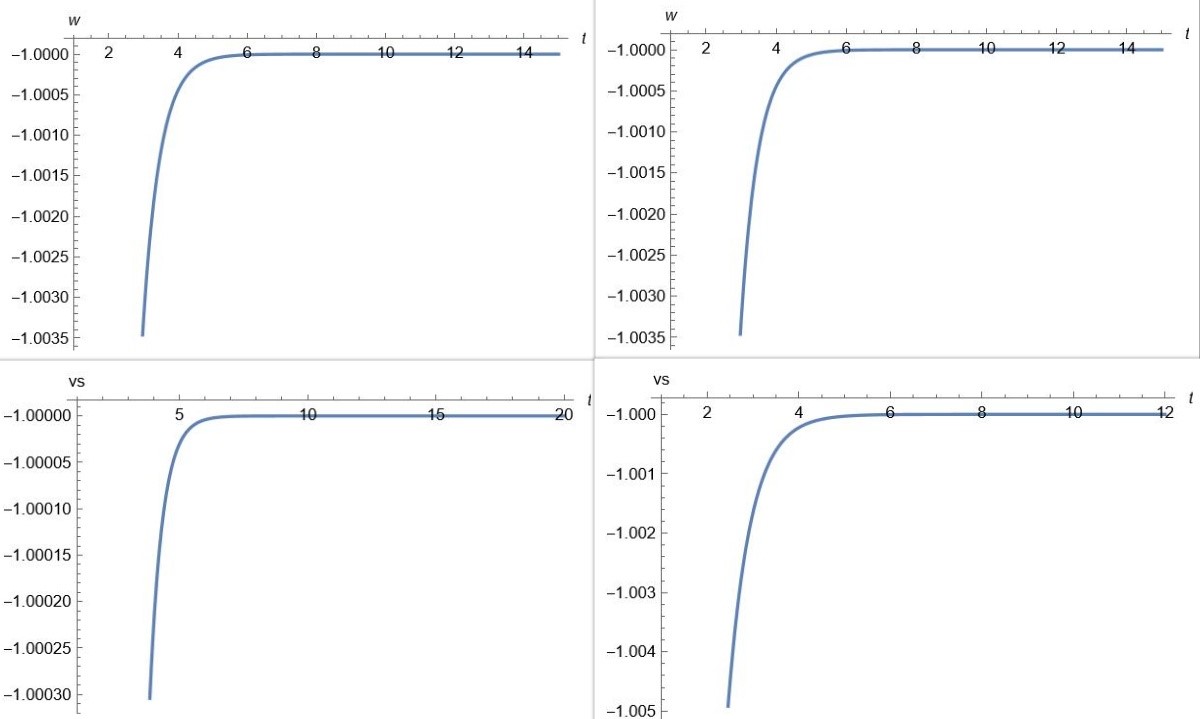}
    \caption{Plots of the dark energy EOS parameter and squared sound speed for first LR ansatz, for both linear and nonlinear regimes for the Hubble horizon case. The upper panel shows the EOS parameter and the lower panel shows the squared sound speed, going from linear to nonlinear regimes from left to right}
    \label{hubble horizon lr normal}
\end{figure}
In figure \ref{hubble horizon lr normal} we have plotted the EOS parameter and squared sound speed for both linear and nonlinear interactions in the ansatz \eqref{lr1} \footnote{For all the plots, here and thereafter, we have taken the following values for all the parameters in all the 6 ansatz mentioned above : $\Omega_{d} =0.69$, $\Omega_{m}=0.258$, $ \alpha =2 $, $ a_{0}=1 $,$a_{f}=1$, $ b=0.03$, $ \beta =2 $, $ c=0.4$, $H_{0}=1$, $H_{1}= 0.8$, $\lambda =2$ and $\eta_{0}=4$. These parameters have been taken with the above values to be consistent with observational bounds and also with theoretical requirements on the same.}. As is seen in figure \ref{hubble horizon lr normal}, the EOS parameter for the first ansatz in the Hubble horizon cutoff is hovering around realistic values and asymptotes to -1 as in the case of usual rips. However, the squared sound speed is always negative and so there is no classical stability for the scenario. We also note that the results are the same for both linear and nonlinear regimes, which is a theme we observed for all the cases which we shall discuss from here onwards. Thereby we have only shown the plots for the linear interactions in the below cases as the plots have been seen to be almost the same in both linear and nonlinear regimes in all cases.
\begin{figure}[!ht]
    \centering
    \includegraphics[width=1\linewidth]{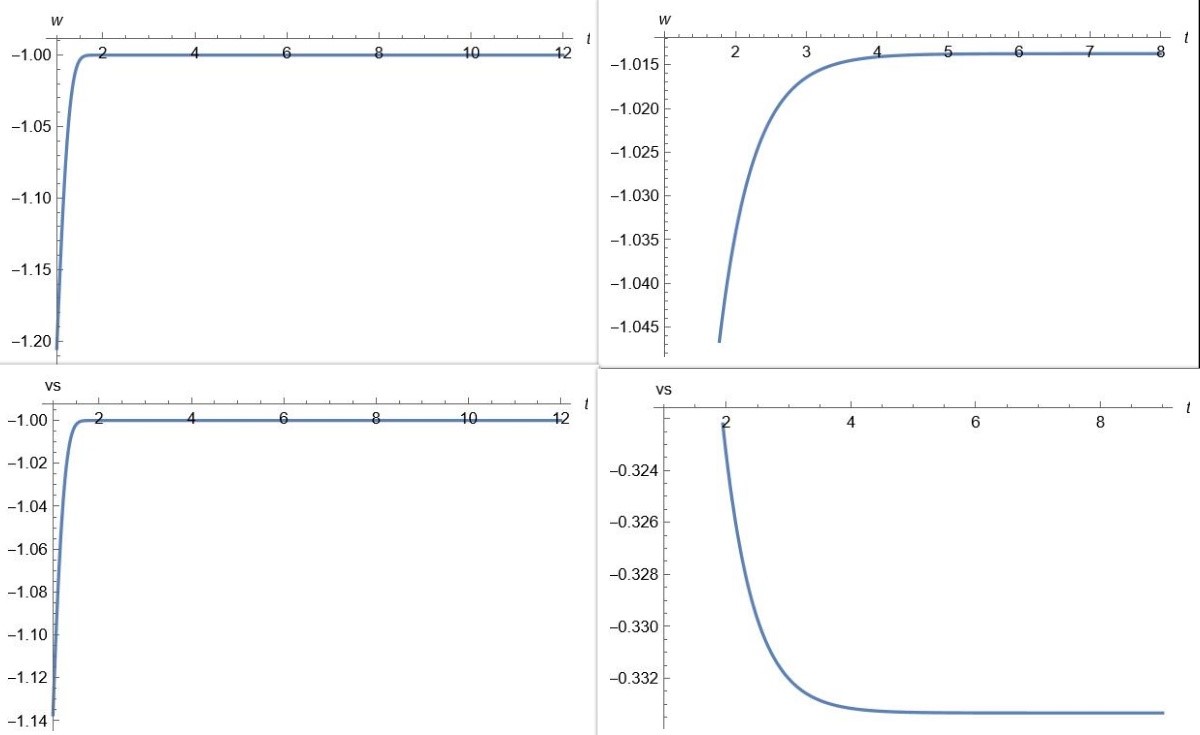}
    \caption{Plots of the dark energy EOS parameter and squared sound speed for second LR ansatz and PR. The upper panel shows the EOS parameter and the lower panel shows the squared sound speed, going from LR to PR from left to right in the linear interaction regime}
    \label{hubble horizon normal lr2 pr }
\end{figure}
In figure \ref{hubble horizon normal lr2 pr } we have plotted for the second LR ansatz \eqref{lr2} and PR \eqref{pr}. What is seen again is that the EOS parameter asymptotes to -1 in both cases and the evolution of the EOS parameter is around acceptable values but the squared sound speed is again completely negative. This trend stays the same for the case of the nonlinear case too, which is why we haven't shown the plot for that case.
\begin{figure}[!ht]
    \centering
    \includegraphics[width=1\linewidth]{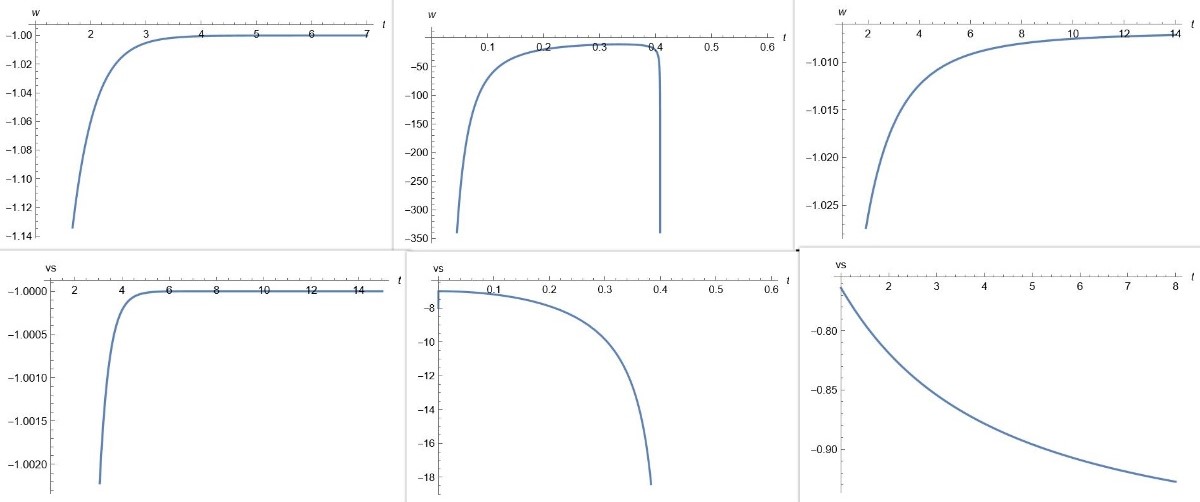}
    \caption{Plots of the dark energy EOS parameter and squared sound speed for LR, Big Freeze and Asymptotic dS sceanrios in the Braneworld for the Hubble horizon case. The upper panel shows the EOS parameter and the lower panel shows the squared sound speed, going from LR to Big Freeze to Asymptotic dS from left to right, in the linear interaction regime}
    \label{hubble horizon normal braneworld }
\end{figure}
In figure \ref{hubble horizon normal braneworld } we have now plotted for the EOS parameter and squared sound speed in LR \eqref{blr}, Big Freeze \eqref{bf} and asymptotic dS \eqref{ads} again only in the linear regime for reasons mentioned above. We see that the Little rip and Asymptotic dS cases show acceptable evolution of the w parameter but peculiar behaviour is seen in the case of Big Freeze. In all these cases, again, there persists classical instability. From here we can conclude that none of the rip or rip-like events can be viable in the Hubble horizon cutoff given the conventional DE model. 
\begin{figure}[!ht]
    \centering
    \includegraphics[width=1\linewidth]{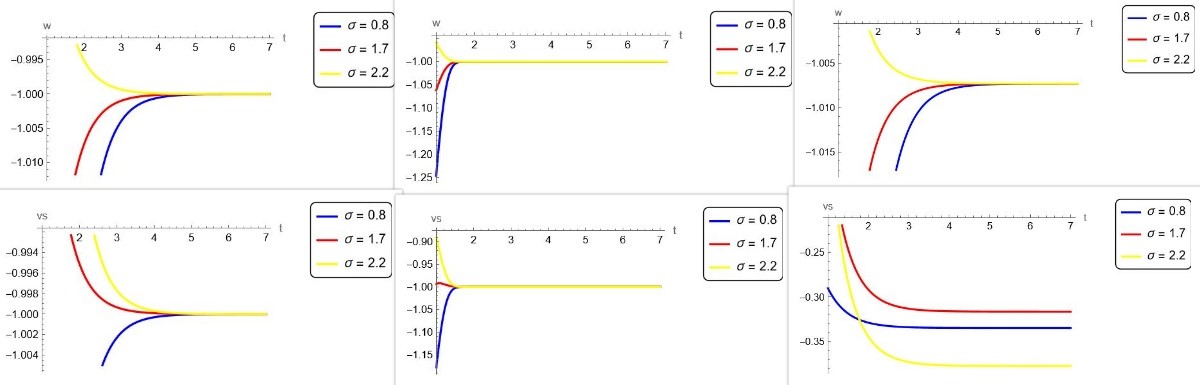}
    \caption{Plots of the dark energy EOS parameter and squared sound speed for both LR ansatz and PR in the Tsallis paradigm for various values of the Tsallis parameter for the Hubble horizon case. The upper panel shows the EOS parameter and the lower panel shows the squared sound speed, going from LR 1 to LR 2 to PR from left to right, in the linear interaction regime}
    \label{hubble horizon tsallis lr pr }
\end{figure}
In figure \ref{hubble horizon tsallis lr pr } we have plotted for the EOS parameter and squared sound speed in both LR ansatz and PR in the case of the Tsallis paradigm. We have plotted for the whole spectrum of the Tsallis parameter values to tack in generality. We see that the EOS parameter again asymptotes around -1 and has acceptable behaviour but even the Tsallis case is riddled with instabilities for all ranges of the free parameter. 
\begin{figure}[!ht]
    \centering
    \includegraphics[width=1\linewidth]{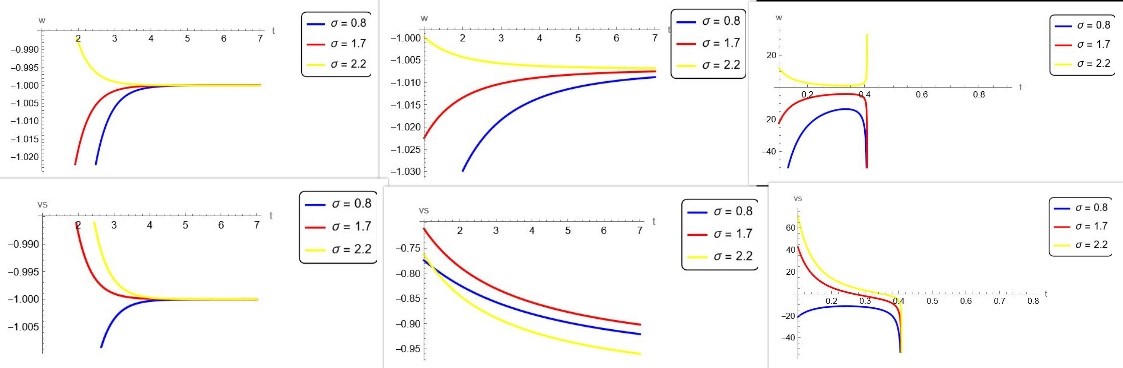}
    \caption{Plots of the dark energy EOS parameter and squared sound speed for LR, Big Freeze and Asymptotic dS scenarios in the Braneworld in the Tsallis case for the Hubble horizon case. The upper panel shows the EOS parameter and the lower panel shows the squared sound speed, going from LR to Asymptotic dS to Big Freeze from left to right, in the linear interaction regime}
    \label{hubble horizon tsallis braneworld }
\end{figure}
In figure \ref{hubble horizon tsallis braneworld } we have plotted similarly for the Tsallis case in the Braneworld for LR, Big Freeze and asymptotic dS. We again see that there is some peculiar evolution in the case of the big freeze in its EOS but the EOS evolution for LR and asymptotic dS seem fine. Again there are instabilities, even though the Big Freeze case starts off slightly positive with regards to its squared sound speed but becomes negative again.
\begin{figure}[!ht]
    \centering
    \includegraphics[width=1\linewidth]{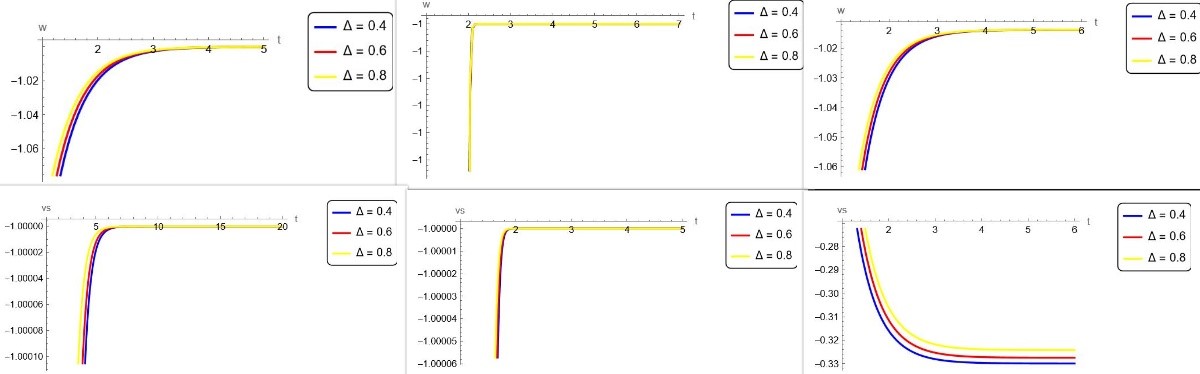}
    \caption{Plots of the dark energy EOS parameter and squared sound speed for both LR ansatz and PR in the Tsallis paradigm for various values of the Tsallis parameter for the Hubble horizon case. The upper panel shows the EOS parameter and the lower panel shows the squared sound speed, going from LR 1 to LR 2 to PR from left to right, in the linear interaction regime}
    \label{hubble horizon barrow non braneworld }
\end{figure}
We then plotted for the same in figure \ref{hubble horizon barrow non braneworld } for the Barrow model with for the both LR ansatz and PR. We again see that the EOS parameter has acceptable evolution but the squared sound speed is negative in all cases which again shows instabilities.  
\begin{figure}[!ht]
    \centering
    \includegraphics[width=1\linewidth]{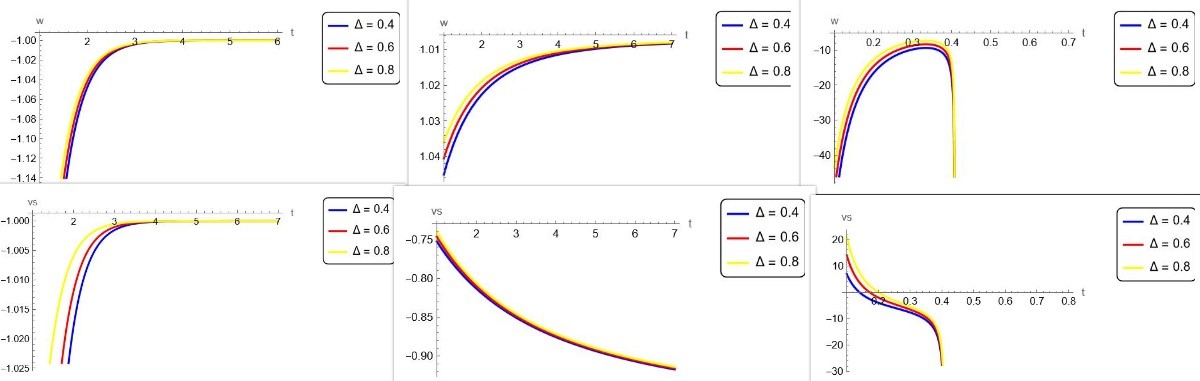}
    \caption{Plots of the dark energy EOS parameter and squared sound speed for LR, Big Freeze and Asymptotic dS scenarios in the Braneworld in the Barrow case for the Hubble horizon case. The upper panel shows the EOS parameter and the lower panel shows the squared sound speed, going from LR to Asymptotic dS to Big Freeze from left to right, in the linear interaction regime}
    \label{hubble horizon barrow braneworld}
\end{figure}
In the case of the Barrow model with a Braneworld, we have plotted for the LR, Big Freeze and Asymptotic dS cases in figure \ref{hubble horizon barrow braneworld}. We see that the trend is similar to the usual non Barrow case even for all ranges of the deformation parameter and again there is no stability. So we would like to conclude from here that these rip and rip-like events are not viable even in Tsallis and Barrow in the Hubble horizon cutoff. 
\section{Future events with particle horizon cutoff}
In the case of the particle horizon cutoff, we have \begin{equation*}
    \frac{\dot{L}}{L} = \left( H + \frac{1}{L} \right)  
\end{equation*}
Now using this, we can write \begin{equation}
    \dot{\rho} = -2 \rho H \left( 1 + \frac{ \sqrt{\Omega_{d}}}{c} \right) 
\end{equation}
From this expression itself we can come to a very strong conclusion. For any rip event, either big rip, little rip, pseudo rip etc. we need the energy denstiy of DE to be monotonically increasing but in this case it is completely evident that the energy density is monotonically decreasing. Hence in the particle horizon cutoff with the usual DE model, we cannot have any rip scenario occurring. Similarly in the case of the Barrow model with the particle horizon cutoff we have \begin{equation}
    \dot{\rho} = - (2-\Delta) \rho H \left( 1 + \frac{\sqrt{\Omega_{d}}}{c} \right)
\end{equation}
We again see that the energy density will be monotonically decreasing in this case as well, as $\Delta \leq 1$. In the case of the Tsallis model we have \begin{equation}
    \dot{\rho} = - (4-2 \sigma) \rho H \left( 1 + \frac{\sqrt{\Omega_{d}}}{c} \right)
\end{equation}
In the case of the Tsallis model there could be a possibility energy density could be monotonically increasing when $\sigma > 2$ and so rips may be possible. 
\begin{figure}[!ht]
    \centering
    \includegraphics[width=1\linewidth]{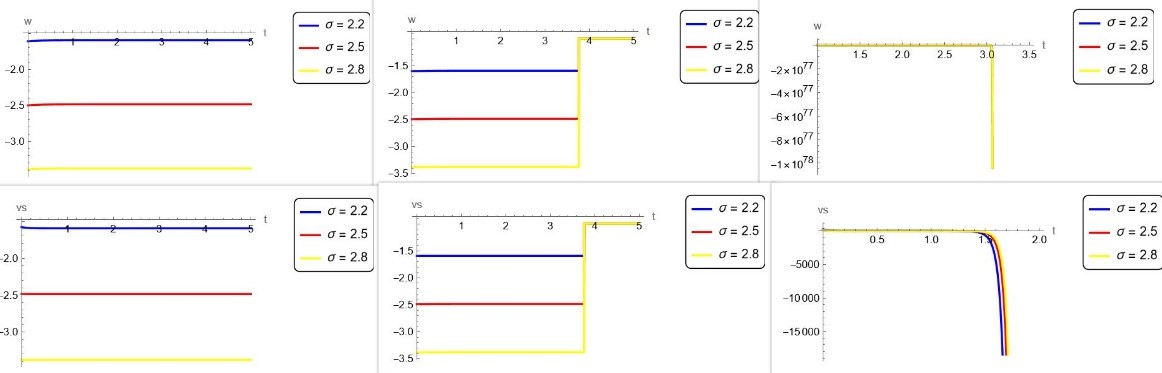}
    \caption{Plots of the dark energy EOS parameter and squared sound speed for both LR ansatz and for PR with particle horizon case. The upper panel shows the EOS parameter and the lower panel shows the squared sound speed, going from linear to nonlinear regimes from left to right}
    \label{Tsallis non braneworld}
\end{figure}
In figure \ref{Tsallis non braneworld} we have plotted for both LR ansatz and PR as we had discussed before, but for Tsallis models with particle horizon cutoff with $\sigma > 2$. Again, there is no difference in the plots and their conclusions in the linear and non-linear interaction cases. We see that similar to the cases as of before, the squared sound speed remains negative in all cases, while for larger values of $\sigma$, even the equation of state parameter touches an unrealistic range for LR as well, while it is certainly true for all range in PR. 
\begin{figure}[!ht]
    \centering
    \includegraphics[width=1\linewidth]{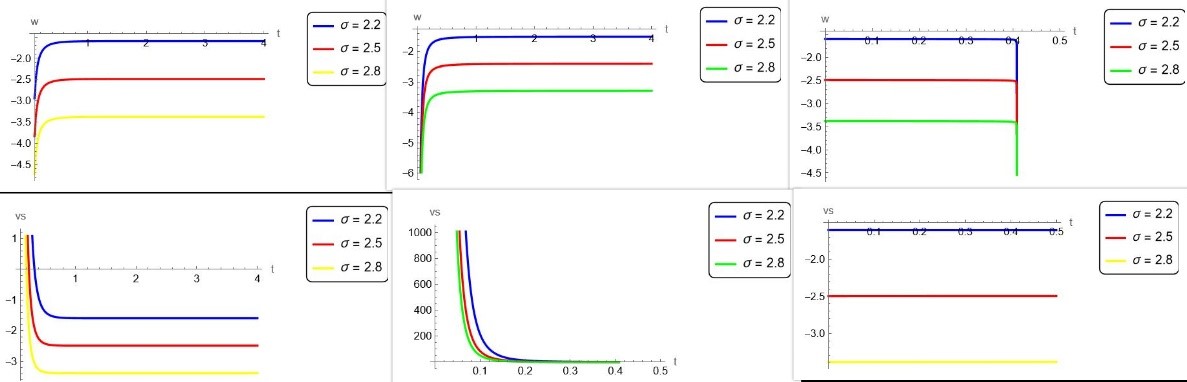}
    \caption{Plots of the dark energy EOS parameter and squared sound speed for both LR ansatz and for PR with particle horizon case. The upper panel shows the EOS parameter and the lower panel shows the squared sound speed, going from linear to nonlinear regimes from left to right}
    \label{Tsallis braneworld}
\end{figure}
In figure \ref{Tsallis braneworld} we have plotted for the LR, asmyptotic dS and Big Freeze cases in a BraneWE again see that there is not much hope for classical stability even in these cases, while the EOS parameter is fine lower values of $\sigma$ for LR and asymptotic dS case, for the Big Freeze even that is beyond an acceptable range. 
\\
\\
\section{Future events with event horizon cutoff}
\begin{figure}[!ht]
    \centering
    \includegraphics[width=1\linewidth]{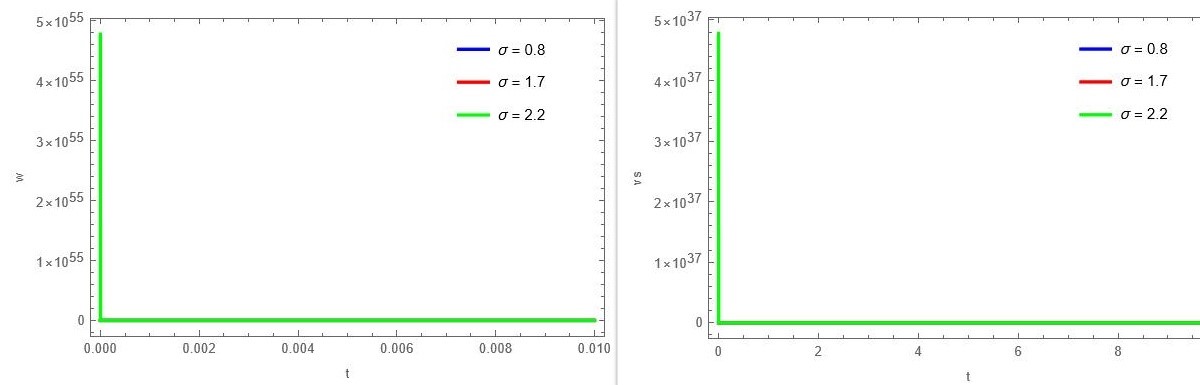}
    \caption{Plots of dark energy EOS parameter and squared sound speed against time in the non-linear interaction for the Braneworld Little rip case in the Tsallis scenario}
    \label{lrq2}
\end{figure}

\begin{figure}[!ht]
    \centering
    \includegraphics[width=1\linewidth]{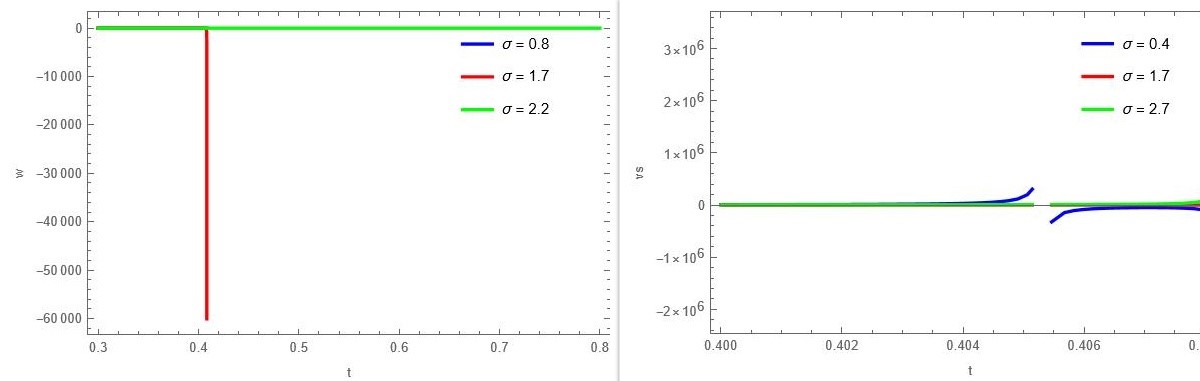}
    \caption{Plots of dark energy EOS parameter and squared sound speed against time in the linear interaction for the Braneworld Big Freeze case in the Tsallis scenario}
    \label{bfq1}
\end{figure}

\begin{figure}[!ht]
    \centering
    \includegraphics[width=1\linewidth]{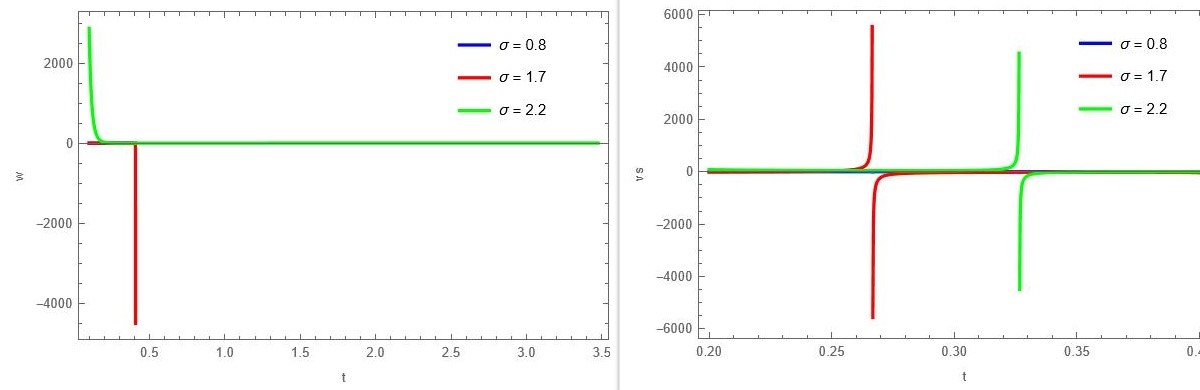}
    \caption{Plots of dark energy EOS parameter and squared sound speed against time in the non-linear interaction for the Braneworld Big Freeze case in the Tsallis scenario}
    \label{bfq2}
\end{figure}

\begin{figure}[!ht]
    \centering
    \includegraphics[width=1\linewidth]{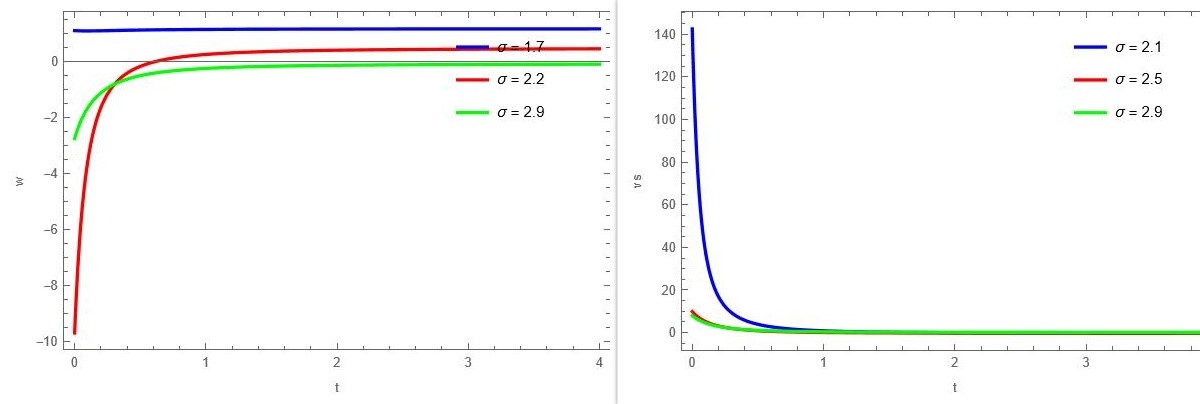}
    \caption{Plots of dark energy EOS parameter and squared sound speed against time in the linear interaction for the Braneworld Asymptotic dS case in the Tsallis scenario}
    \label{adsq1}
\end{figure}

\begin{figure}[!ht]
    \centering
    \includegraphics[width=1\linewidth]{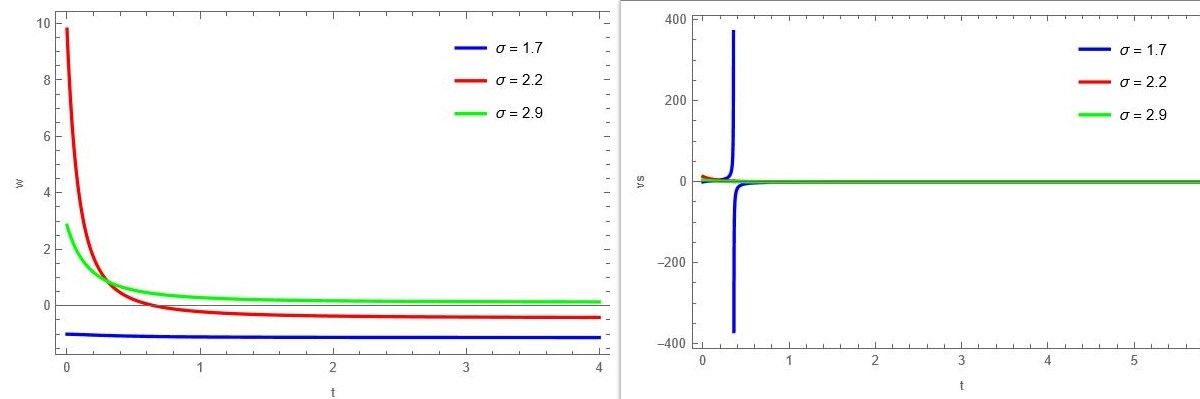}
    \caption{Plots of dark energy EOS parameter and squared sound speed against time in the non-linear interaction for the Braneworld Asymptotic dS case in the Tsallis scenario}
    \label{adsq2}
\end{figure}

In the case of the event horizon, we have the following in the usual HDE model
\begin{equation}
\label{rnormaleh}
    \dot{\rho} = 2 \rho \left( \left( \frac{\Omega_{d} H^2}{c^2}\right)^{\frac{1}{2}} - H  \right) 
\end{equation}
The EOS parameter is
\begin{equation} \label{wnormaleh}
    w= -1 - \frac{1}{3 H \Omega_{d}} \left[ (2) \Omega_{d} \left( \left( \frac{\Omega_{d} H^2}{c^2}\right)^{\frac{1}{2}} - H \right) - \frac{Q}{3 H^2} \right]
\end{equation}
While in the Tsallis model \begin{equation} \label{rtsaeh}
    \dot{\rho} = (4 - 2 \sigma) \rho \left( \left( \frac{\Omega_{d} H^2}{c^2}\right)^{\frac{1}{4 - 2 \sigma}} - H  \right) 
\end{equation}
\begin{equation} \label{wtsaeh}
   w= -1 - \frac{1}{3 H \Omega_{d}} \left[ (4 - 2 \sigma) \Omega_{d} \left( \left( \frac{\Omega_{d} H^2}{c^2}\right)^{\frac{1}{4 - 2 \sigma}} - H \right) - \frac{Q}{3 H^2} \right]
\end{equation} 
And for the Barrow model, we have \begin{equation} \label{rbareh}
    \dot{\rho} = (2-\Delta) \rho \left( \left( \frac{\Omega_{d} H^2}{c^2}\right)^{\frac{1}{2-\Delta}} - H  \right) 
\end{equation}
\begin{equation} \label{wbareh}
    w= -1 - \frac{1}{3 H \Omega_{d}} \left[ (2-\Delta) \Omega_{d} \left( \left( \frac{\Omega_{d} H^2}{c^2}\right)^{\frac{1}{2-\Delta}} - H \right) - \frac{Q}{3 H^2} \right]
\end{equation}
\begin{figure}[!ht]
    \centering
    \includegraphics[width=1\linewidth]{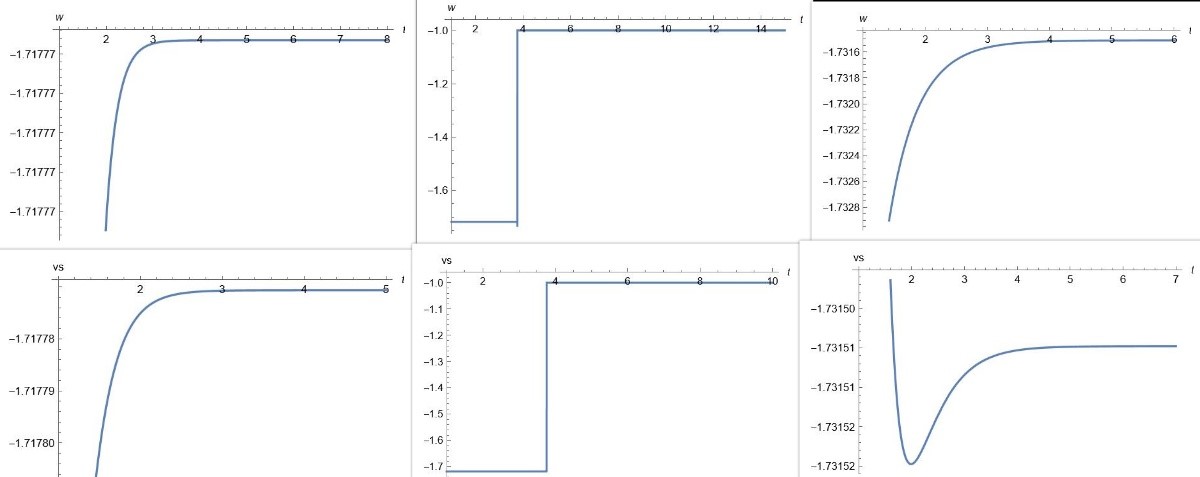}
    \caption{Plots of the dark energy EOS parameter and squared sound speed for both LR ansatz and PR. The upper panel shows the EOS parameter and the lower panel shows the squared sound speed, going from LR 1 to LR 2 to PR from left to right, in the linear interaction regime}
    \label{eh normal}
\end{figure}
In figure \ref{eh normal}  we have plotted for both LR ansatz and PR for the case of the event horizon cutoff in the conventional HDE model. What one sees is a recurring pattern again, namely that the squared sound speed is negative while the EOS parameter hovers around viable values. 
\begin{figure}[!ht]
    \centering
    \includegraphics[width=1\linewidth]{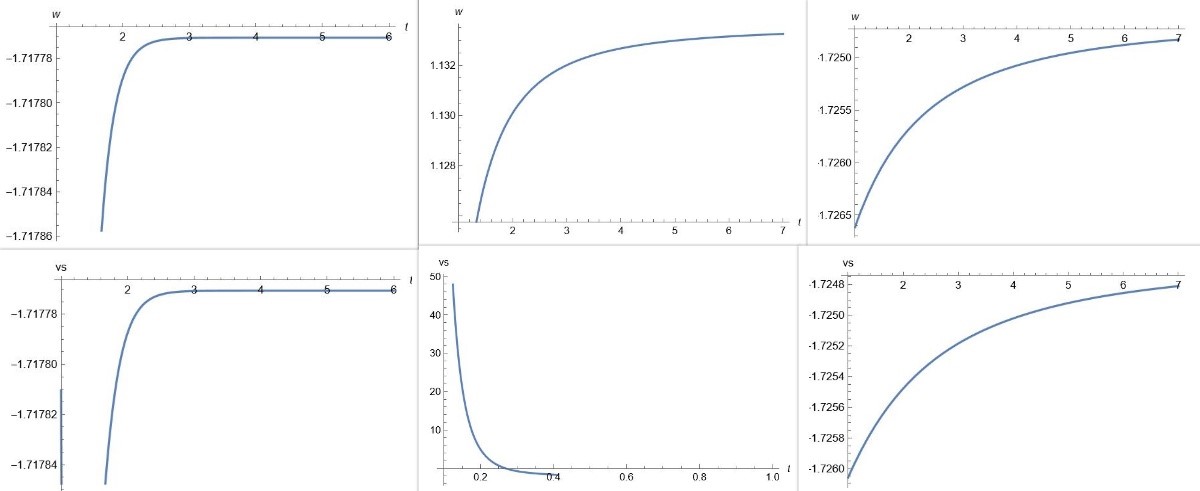}
    \caption{Plots of the dark energy EOS parameter and squared sound speed for LR, Big Freeze and asymptotic dS. The upper panel shows the EOS parameter and the lower panel shows the squared sound speed, going from LR to Big Freeze to asymptotic dS from left to right, in the linear interaction regime}
    \label{eh brane}
\end{figure}
In figure \ref{eh brane} we have plotted for the Braneworld case for the same and we again see that there is not much hope for viable scenarios. The Big Freeze case initially produces positive squared sond speed values which eventually become negative again. 
\begin{figure}[!ht]
    \centering
    \includegraphics[width=1\linewidth]{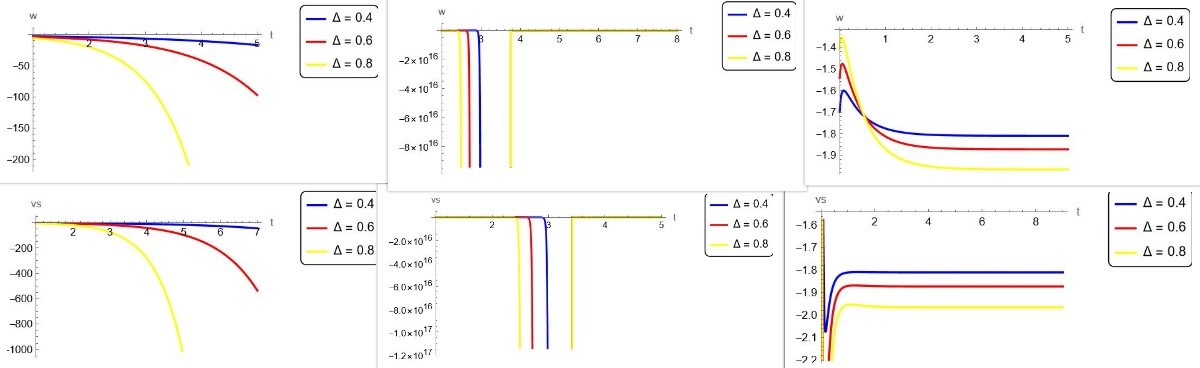}
    \caption{Plots of the dark energy EOS parameter and squared sound speed for both LR ansatz and PR in the Barrow paradigm. The upper panel shows the EOS parameter and the lower panel shows the squared sound speed, going from LR 1 to LR 2 to PR from left to right, in the linear interaction regime}
    \label{eh barrow normal }
\end{figure}
In the Barrow case for both the LR ansatz and PR as in figure \ref{eh barrow normal }, we see that the first LR unrealistic values for both w and the squared sound speed parameter as well. The case of the PR also produces overtly negative values for both the squared sound speed and EOS parameters, while the second LR seemingly gives values of the squared sound speed close to 0 but it is still negative. 
\begin{figure}[!ht]
    \centering
    \includegraphics[width=1\linewidth]{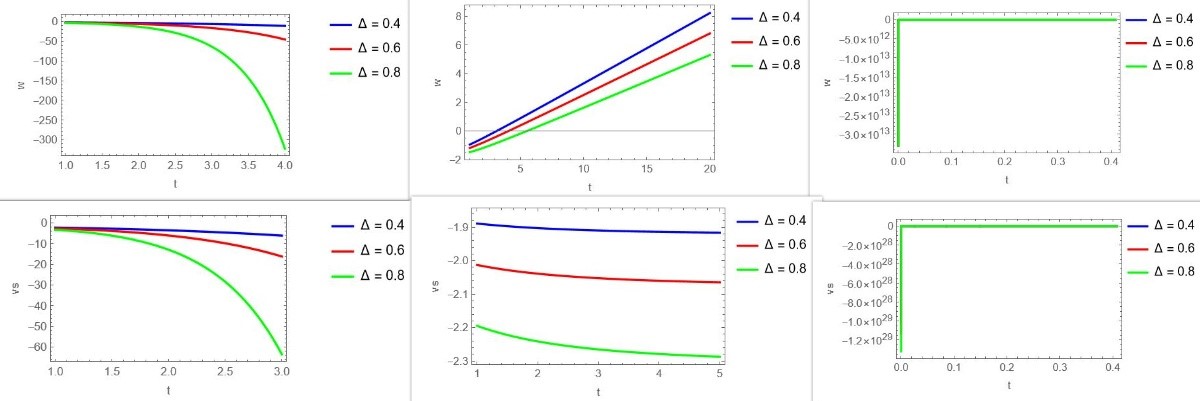}
    \caption{Plots of the dark energy EOS parameter and squared sound speed for LR, Big Freeze and asymptotic dS in the Barrow paradigm. The upper panel shows the EOS parameter and the lower panel shows the squared sound speed, going from LR to Big Freeze to asymptotic dS from left to right, in the linear interaction regime}
    \label{eh barrow brane}
\end{figure}
In figure \ref{eh barrow brane} we see that the Barrow model in the Braneworld also suffers from similar issues for all future events. There is no hope for stability and even the evolution of the dark energy EOS parameter is seemingly going off course for various values of the free parameter. 
\begin{figure}[!ht]
    \centering
    \includegraphics[width=1\linewidth]{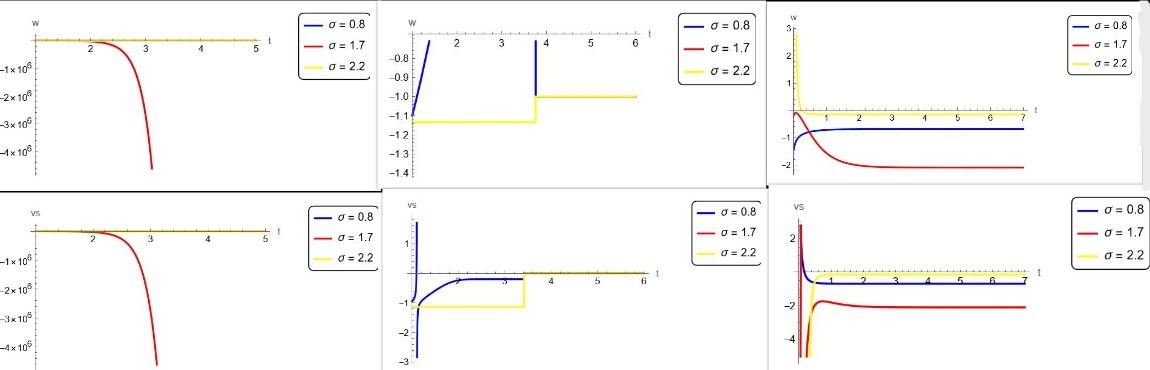}
    \caption{Plots of the dark energy EOS parameter and squared sound speed for both LR ansatz and PR in the Tsallis paradigm. The upper panel shows the EOS parameter and the lower panel shows the squared sound speed, going from LR 1 to LR 2 to PR from left to right, in the linear interaction regime}
    \label{eh tsallis normal }
\end{figure}
In figure \ref{eh tsallis normal } we see that for the case of the first LR variant, there is unrealistic evolution for both the EOS parameter and the squared sound speed parameter as well. The second LR variant shows very erratic behaviour but is closer to acceptable values for the EOS parameter however for the squared sound speed parameter it is still negative. For the PR case the story still remains the same and we again have no scenario where there is even a slight hope of stability as well. 
\begin{figure}[!ht]
    \centering
    \includegraphics[width=1\linewidth]{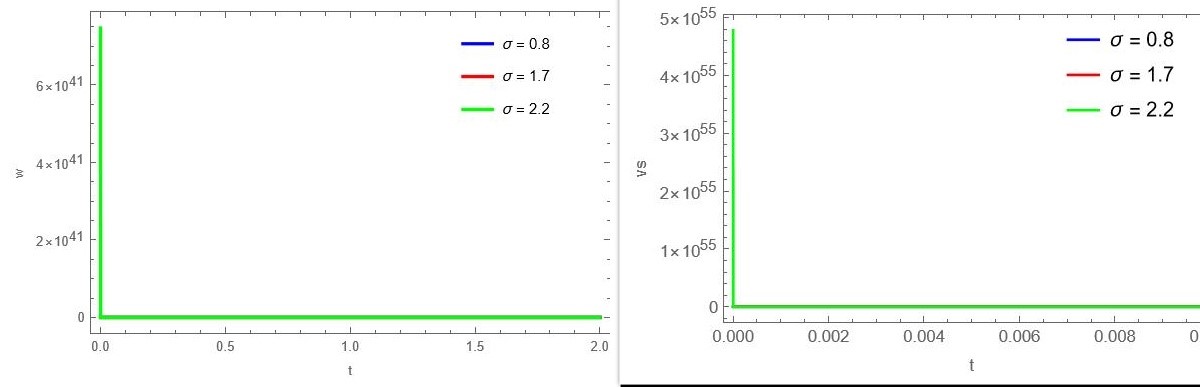}
    \caption{Plots of dark energy EOS parameter and squared sound speed against time in the linear interaction for the Braneworld Little rip case in the Tsallis scenario}
    \label{lrq1}
\end{figure}
In figure \ref{lrq1}, we have plotted the dark energy EOS parameter and the squared sound speed for various values of the Tsallis parameter in the linear interaction for the Little rip scenario and we clearly see that while the squared sound speed seems to be in acceptable region ($\sim 0$ ), the EOS parameter taking values $\sim 0$ is quite alarming and shows unrealistic values of the parameter. Exactly similar trend is seen when one plots for the same scenario in the non-linear model, as seen in figure \ref{lrq2}.
When one discusses the Big Freeze singularity in the linear interaction as in figure \ref{bfq1}, we see again that the EOS parameter takes values $\sim 0$ while squared sound speed shows values slightly below 0 or $\sim 0$ for various ranges of the Tsallis parameter.  
Discussing the same in the non linear interaction in figure \ref{bfq2}, we see that the EOS parameter is still acting the same but the squared sound speed now shows jumps and takes on absurdly negative or positive values, which are both not acceptable.
When one discusses the asymptotic dS case however, we do see some viable results. As plotted for the linear interaction case in figure \ref{adsq1}, we see that for high values of $\sigma$, preferrably $\sigma > 2.5$, the EOS parameter is in acceptable regions hovering around realisitc negative values while at the same time, the squared sound speed maintains very small positive values close to 0.  
When one considers the non-linear interaction in this scenario as in figure \ref{adsq2}, we see that smaller values of $\sigma$, in the region $1.5 < \sigma < 2$ produce viable values of the EOS parameter but all models for all ranges of $\sigma$ do not lead to viable sound speeds as they hover around the value of -1 only.
So one sees that from all these rip scenarios available, the only one which is looking viable as of now is the Asmyptotic dS scenario with linear interaction for high values of the Tsallis parameter. An interesting thing to note in the Tsallis cases is that there is at least some degree of difference in behaviours in linear and nonlinear interaction regimes in the behaviour of some of the scenarios.
\\
\\
\section{Thermodynamics and energy conditions}
It is also important to address another aspect that could help ensure the consistency of holographic dark energy models, namely thermodynamics. The thermodynamic implications, especially the validity of holographic dark energy models concerning the generalized second law of thermodynamics, were recently examined in \cite{Mamon:2020wnh}. It is well known that the thermodynamic analysis of gravitational theories offers a fascinating interface with cosmology. The thermodynamic properties that hold for a black hole are equally applicable to a cosmological horizon. Moreover, the first law of thermodynamics, which holds at a black hole horizon, can also be derived from the first Friedmann equation in the FLRW universe when the universe is bounded by an apparent horizon. This provides strong motivation to select the apparent horizon as the cosmological horizon to examine the thermodynamic properties of any cosmological model. 

Motivated by these arguments, we consider the universe as a thermodynamic system bounded by the cosmological apparent horizon, with the radius \cite{Bak:1999hd} given by:

\begin{equation} 
    r_{h} = \left( H^2 + \frac{k}{a^2} \right)^{-1/2}
\end{equation}

For a flat universe $(k=0)$, this simplifies to:

\begin{equation} \label{rh}
    r_{h} = \frac{1}{H}
\end{equation}

Considering that the major contributions to the entropy of the universe come from the dark sectors in addition to the cosmological horizon, we can write:

\begin{equation}
    S_{tot} = S_{\Lambda} + S_{m} + S_{h}
\end{equation}

where $S_{h}$ is the horizon entropy. The time evolution of $S_{\Lambda}$ and $S_{m}$ can be expressed using the first law of thermodynamics as follows:

\begin{equation} \label{firstlawd}
    T \dot{S}_{\Lambda} = \dot{E}_{\Lambda} + p_{\Lambda} \dot{V}
\end{equation}

\begin{equation} \label{firstlawm}
    T \dot{S}_{m} = \dot{E}_{m} + p_{m} \dot{V}
\end{equation}

where $V = \frac{4 \pi r_{h}^3}{3}$ is the horizon volume, with 

$$E_{\Lambda}= \frac{4 \pi r_{h}^3 \rho_{\Lambda}}{3}$$ 

and 

$$E_{m}= \frac{4 \pi r_{h}^3 \rho_{m}}{3}$$

The time evolutions of the energy densities are given by the continuity equations \eqref{contd} and \eqref{contm}. The temperature \( T \) of the apparent horizon is given by \footnote{We have considered the Gibbons-Hawking temperature here for the cosmological apparent horizon. This has been a consideration which has been prevalent in studies of the thermodynamics of HDE before as well, see for example \cite{Mamon:2020wnh,Chakraborty:2021uzp}. This temperature is indeed apt for a dS epoch and this consideration is fine in our study as we are interested in the properties of the very late universe here, which evolves towards an dS space with equilibrium conditions intact.}
\begin{equation}
    T = \frac{1}{2 \pi r_{h}} = \frac{H}{2 \pi}
\end{equation}  
The entropy of the horizon depends on the HDE model we consider. In the case of the simple HDE, we use the usual Bekenstein-Hawking entropy, which can be written as:

\begin{equation} 
    S_{h} = \frac{A}{A_{0}}  
\end{equation} 

where \( A_{0} = \frac{1}{4 G} \) is the Planck area, and we can rewrite this as:

\begin{equation} \label{sbenk}
    S_{h} = \gamma r_{h}^2
\end{equation}

Here, \( \gamma \) is a positive constant (in this case equal to \( \pi / G \), but the utility of this expression for \( S_{h} \) will soon be clear). When considering Barrow entropy, we have:

\begin{equation}
     S_{h} = \left( \frac{A}{A_{0}} \right)^{1 + \Delta / 2}
 \end{equation}

Using the horizon radius, this can be expressed as:

\begin{equation} \label{sbar}
     S_{h} = \gamma_{b} r_{h}^{2 + \Delta}
 \end{equation}

where 

$$\gamma_{b} = \left( \frac{4 \pi}{A_{0}} \right)^{1 + \Delta / 2}$$ 

is a positive constant. For the Tsallis HDE scenario, the horizon entropy is given by:

\begin{equation} \label{stsa}
    S_{h} = \gamma_{t} r_{h}^{2 \sigma}
\end{equation}

where \( \gamma_{t} \) is again a constant in terms of the Planck area and the Tsallis parameter. We can write the derivative of the total entropy of the universe in the case of the Hubble horizon cutoff in the conventional HDE model as \begin{equation} \label{hubbles1}
    \dot{S} = \frac{2 \pi}{H} \Bigg[ \frac{8 \pi \dot{H} \Omega_{d}}{H} - \frac{24 \pi \dot{H}}{H} - 4 \pi \left( \frac{Q}{3 H^2} + 3 H \Omega_{m} (1+ w_{m}) \right) \Bigg] - \frac{2 \gamma \dot{H}}{H^3}
\end{equation}
Similarly we can write for the Barrow and Tsallis cases as \begin{equation} \label{hubblesb}
    \dot{S} = \frac{2 \pi}{H} \Bigg[ \frac{4 \pi \dot{H} \Omega_{d} (2- \Delta)}{H} - \frac{24 \pi \dot{H}}{H} - 4 \pi \left( \frac{Q}{3 H^2} + 3 H \Omega_{m} (1+ w_{m}) \right) \Bigg] - \frac{(2 + \Delta) \gamma_{b} \dot{H}}{H^2 H^{1 + \Delta }}
\end{equation}
\begin{equation} \label{hubblest}
    \dot{S} = \frac{2 \pi}{H} \Bigg[ \frac{4 \pi \dot{H} \Omega_{d} (4 - 2 \sigma)}{H} - \frac{24 \pi \dot{H}}{H} - 4 \pi \left( \frac{Q}{3 H^2} + 3 H \Omega_{m} (1+ w_{m}) \right) \Bigg] - \frac{2 \gamma_{t} \sigma \dot{H}}{H^2 H^{2 \sigma - 1 }}
\end{equation}
Similarly we can find the entropy for the event horizon cases for the conventional, Tsallis and Barrow models as \begin{equation} \label{sevent1}
    \dot{S} = \frac{2 \pi}{H} \Bigg[ 8 \pi  \Omega_{d} \left( \frac{\sqrt{\Omega_{d}}}{c} - 1 \right) - \frac{24 \pi \dot{H}}{H} - 4 \pi \left( \frac{Q}{3 H^2} + 3 H \Omega_{m} (1+ w_{m}) \right) \Bigg] - \frac{2 \gamma \dot{H}}{H^3}
\end{equation}
\begin{equation} \label{seventt}
    \dot{S} = \frac{2 \pi}{H} \Bigg[ \frac{4 \pi \Omega_{d}}{H} (4-2 \sigma) \left( \left( \frac{\Omega_{d} H^2}{c^2} \right)^{(\frac{1}{4-2\sigma})} - H \right) - \frac{24 \pi \dot{H}}{H} - 4 \pi \left( \frac{Q}{3 H^2} + 3 H \Omega_{m} (1+ w_{m}) \right) \Bigg] - \frac{2 \gamma_{t} \sigma \dot{H}}{H^2 H^{2 \sigma - 1 }}
\end{equation}
\begin{equation} \label{seventb}
    \dot{S} = \frac{2 \pi}{H} \Bigg[ \frac{4 \pi \Omega_{d}}{H} (2- \Delta) \left( \left( \frac{\Omega_{d} H^2}{c^2} \right)^{(\frac{1}{2- \Delta})} - H \right) - \frac{24 \pi \dot{H}}{H} - 4 \pi \left( \frac{Q}{3 H^2} + 3 H \Omega_{m} (1+ w_{m}) \right) \Bigg] - \frac{(2 + \Delta) \gamma_{b} \dot{H}}{H^2 H^{1 + \Delta }}
\end{equation}
While we can also find it for the particle horizon case, we will not be seeing the viability of the GSL for this cutoff as we really only have one model (Tsallis with $\sigma>2$ ) to talk about here. We need not plot the evolution of the entropy for every case we have, but we will at least try to get a gist of things here.  
\begin{figure}[!ht]
    \centering
    \includegraphics[width=1\linewidth]{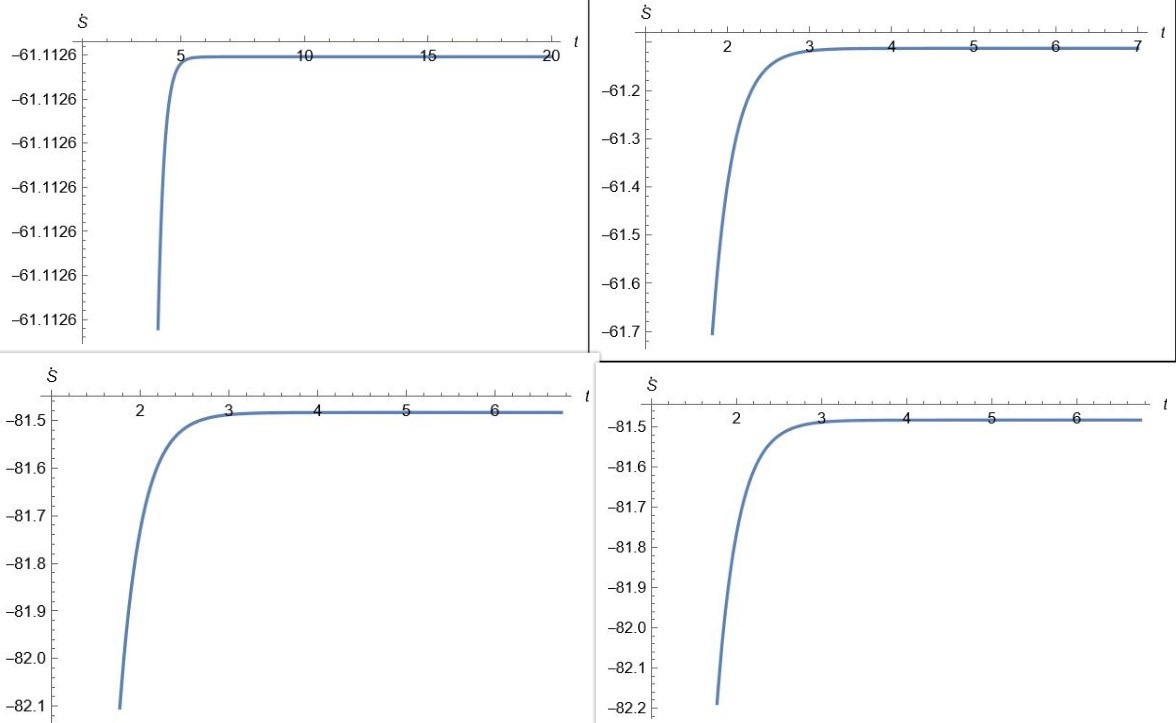}
    \caption{Plots of the derivative of the entropy for the Hubble horizon case with first ansatz of LR. From left to right are the plots for the Hubble and Event horizon cutoffs, with upper panel for $w_{m}=0$ and lower for $w_{m}=1/3$}
    \label{entropys1lr1}
\end{figure}
In figure \ref{entropys1lr1} we have plotted the time derivative of the entropy for both Hubble and event horizon cutoffs and for dust and non-dust forms of dark matter with the first LR ansatz. Note that in the formulations above, the form of dark matter has not been set apriori to simple dust like $w_{m}= 0$  and so it would be interesting to see whether not dust like approximations make a difference too. We see that the interaction schemes or the form of dark matter is not impacting the conclusions in any case, which is conclusively that the generalized second law of thermodynamics is not held in any case. 
\begin{figure}[!ht]
    \centering
    \includegraphics[width=1\linewidth]{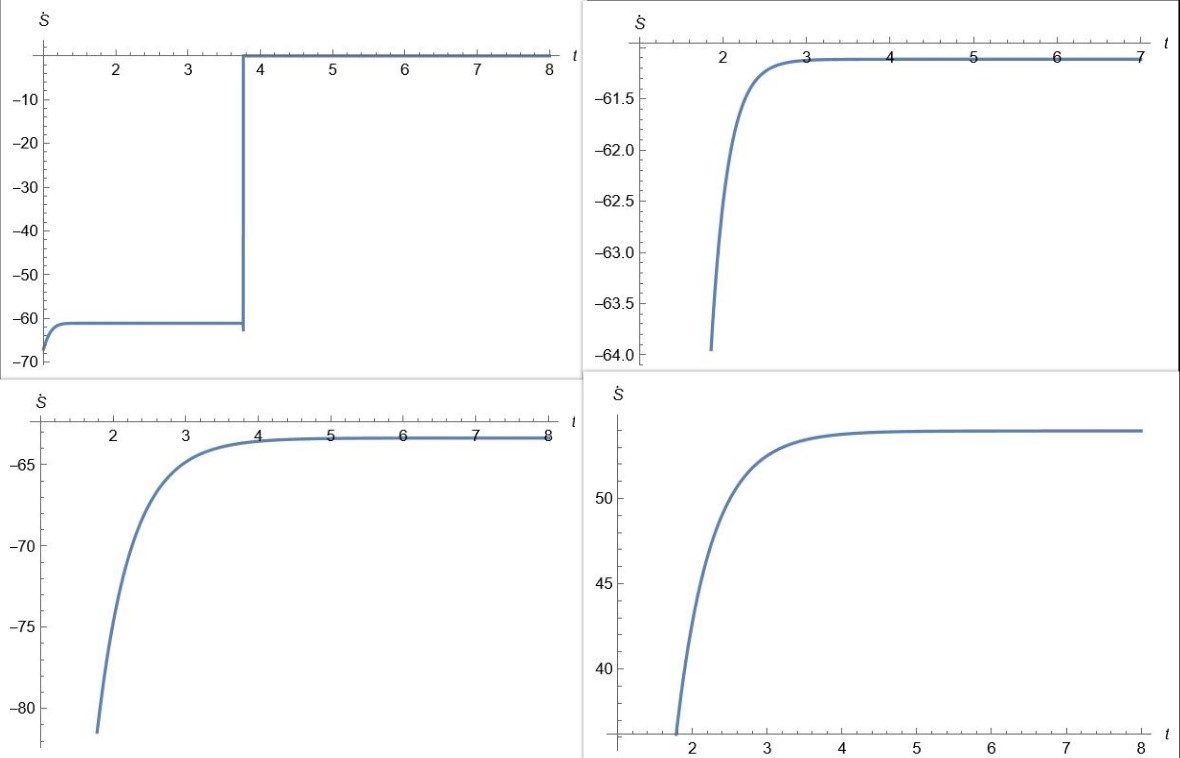}
    \caption{Plots of the derivative of the entropy for the Hubble and event horizon cases for second LR ansatz and PR. From left to right is from Hubble horizon to event horizon cutoff, with upper panel having LR and lower panel PR}
    \label{entropys1lrpr}
\end{figure}
In figure \ref{entropys1lr1} we have plotted for the Hubble and event horizon cutoffs for LR and PR with us again seeing that in the majority of the cases there is no hope for the generalized second law to be satisifed. However, for the PR ansatz with event horizon cutoff ( lower right panel) the GSL is satisfied. 
\begin{figure}[!ht]
    \centering
    \includegraphics[width=1\linewidth]{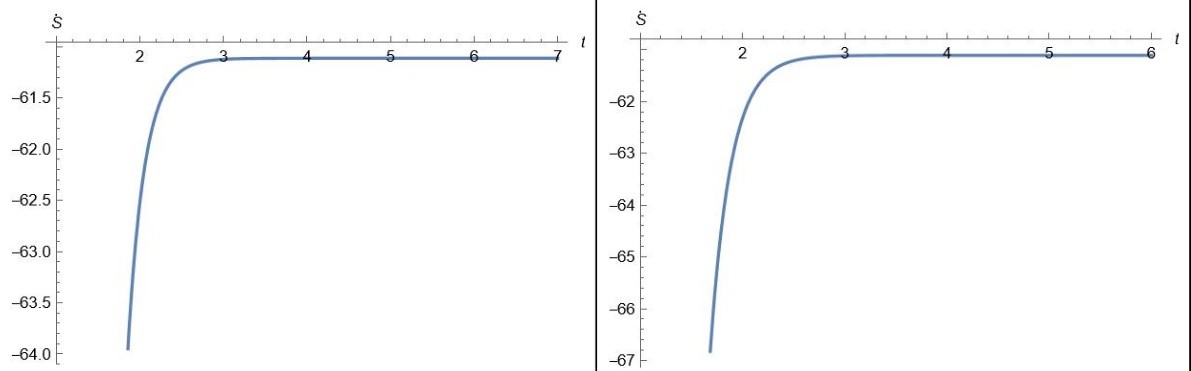}
    \caption{Plots of the derivative of the entropy for the Hubble and event horizon cases for the LR ansatz in the Braneworld}
    \label{entropys1brane}
\end{figure}
We have plotted for the Braneworld case in figure \ref{entropys1brane} with the LR for both the cutoffs and the law is not satisfied. Similar themes were observed for the Big Freeze and Asymptotic dS cases which we have not shown here. 
\begin{figure}[!ht]
    \centering
    \includegraphics[width=1\linewidth]{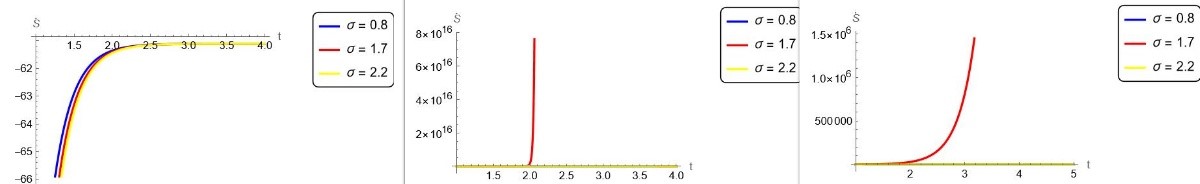}
    \caption{Plots of the derivative of the entropy for the event horizon cutoff for both non-Braneworld and Braneworld ansatz for the Tsallis case, from left to right respectively. }
    \label{entropytsallis}
\end{figure}
\begin{figure}[!ht]
    \centering
    \includegraphics[width=1\linewidth]{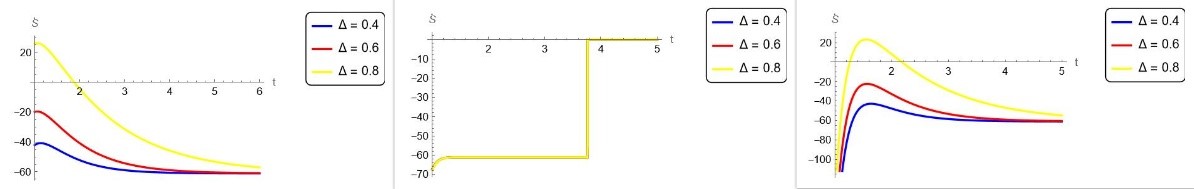}
    \caption{Plots of the derivative of the entropy for the Hubble horizon cutoff for both non-Braneworld and Braneworld ansatz for the Barrow case, from left to right respectively. }
    \label{entropybarrow}
\end{figure}
In figures \ref{entropytsallis} and \ref{entropybarrow} we have plotted the same for both usual LR ansatz and the Braneworld one as well, for the Tsallis case with event horizon and Barrow with Hubble horizon. We see that in the Braneworld and at least one non-Braneworld ansatz, the Tsallis model gives GSL consistent results while the Barrow model with Hubble horizon is not giving any optimistic results. This further reinforces the notion that the Hubble horizon is not well suited at all as a cutoff scheme for the late universe. It is indeed problematic that we are observing negative values for the derivative of the entropy and that is what we wanted to emphasize as well. For example, in figures \ref{entropys1lr1} and \ref{entropys1lrpr}, where one sees negative values of the entropy derivative for cases of Little and Pseudo rip in both standard GR and Braneworld case, the plots are here to further reinforce the idea that these rip scenarios are not very sustainable and not technically possible in these HDE models. Only in very particular parameter regions, for some particular models like one seen in figure \ref{entropytsallis} could one have these rip scenarios to be consistent from the perspective of the second law of thermodynamics. But even when they are consistent from that perspective, they may not be consistent from the perspectives of stability, energy conditions etc. We would also like to note that there has been significant work on cosmological scenarios with such negative entropy evolution, see for example \cite{brevik2013universe}.  
\\
\\
Besides these, another point of technical consistency for cosmological models can pertain to the validity of energy conditions. Four prominent energy conditions in this regard are : \begin{itemize}
    \item null energy condition:  $p + \rho \geq 0$
    \item weak energy condition : $\rho \geq 0$, $\rho + p \geq 0$ 
    \item dominant energy condition : $\rho - p \geq 0$
    \item strong energy condition : $\rho + 3p \geq 0$
\end{itemize}
We would also like to briefly discuss the status quo of these conditions for dark energy sector in the models we have hence considered. In the case of the Hubble horizon cutoff, noting that $\rho (1+ w) = p + \rho$, we can write \begin{equation}
    p + \rho = -H \Bigg[ \frac{Q}{3 H^2} + \frac{2 \dot{H} \Omega_{d}}{H} \Bigg] 
\end{equation} 
From the above equation, it is evident that weak and null energy conditions will be violated for all future events we have discussed here. Similarly, the strong energy condition is also violated, but for the dominant energy condition we have \begin{equation}
    \rho - p = 3 H^2 \Omega_{d} \left( 2 + \frac{1}{3 H \Omega_{d}} \left( \frac{Q}{3 H^2} + \frac{2 \dot{H}\Omega_{d}}{H} \right) \right)
\end{equation}
So the dominant energy condition would hold. Similar conclusions hold true for the Barrow and the Tsallis models with this cutoff as well. Note that there is a catch here, as in the case when the Q term on right hand side of the continuity equation is positive in sign there is a possibility that all energy conditions may be satisfied. This would mean that there is a better possibility of energy conditions being satisfied if the interacting dark sector is one with DE slowly transitioning to DE instead of the other way around as of here. In the case of the particle horizon, we have only considered the Tsallis models with $\sigma> 2$, and in this scenario we can write \begin{equation}
    p + \rho = H \Bigg[ (4-2\sigma) \Omega_{d} \left( H + \left( \frac{\Omega_{d} H^2}{c^2} \right)^{\frac{1}{4-2\sigma}} \right)-\frac{Q}{3 H^2} \Bigg]
\end{equation}
For the case of $\sigma > 2$, we see that the weak and null energy conditions would immediately not be satisfied and similarly not would be the strong energy condition. For the dominant energy condition we have \begin{equation}
    \rho - p = 3 H^2 \Omega_{d} \left( 2 - \frac{1}{3 H\Omega_{d}} (4-2\sigma) \Omega_{d} \left( H + \left( \frac{\Omega_{d} H^2}{c^2} \right)^{\frac{1}{4-2\sigma}} \right)-\frac{Q}{3 H^2}  \right)
\end{equation}
For $\sigma>2$ it is apparent that the dominant energy condition would be satisfied at least, again. For the event horizon case, one can not make such clear statements without plots. If we take the case of only the conventional HDE model with the event horizon cutoff, we have \begin{equation}
    \rho + p = H \Bigg[ 2 \Omega_{d} H \left( 1 - \frac{\sqrt{\Omega_{d}}}{c} \right) - \frac{Q}{3 H^2}  \Bigg]
\end{equation}
\begin{equation}
    \rho - p = 2 + \frac{1}{3 H \Omega_{d}} \Bigg[ 2 \Omega_{d} H \left( 1 - \frac{\sqrt{\Omega_{d}}}{c} \right) - \frac{Q}{3 H^2}  \Bigg]  
\end{equation}
As is clear from the expressions above, in order to make any definitive claims about the status of the energy conditions in this case we would have to plot out these expressions for various choices of the Hubble parameter as per the events we want to consider and we currently do not pursue it here. 
\section{Conclusions}
In this work we have discussed about the intricacies of various possible future scenarios for the universe with holographic dark energy. We started by a brief overview of various holographic dark energy models and cutoffs, alongside brief notes on rips and similar future events. We then firstly discussed about the possibility of accommodating various rip scenarios in the general N-O cutoff for HDEs, showing that the feasibility of all rip scenarios ( or no rip scenario at all) is very easy to realize in the general cutoff scheme. We then turned our attention to the three primitive cutoffs for HDEs, with those being the Hubble, particle and event horizon cutoffs. We discussed in detail about rips in all of these cutoffs, considering the usual HDE model alongside the Tsallis and Barrow models in our work too. We were able to then conclude that in most cases, it is indeed rather rather hard to realize these rip scenarios both from an observational point of view and also from the view of classical stability of these regimes. We finally also discussed about the validity of various energy conditions and the generalized second law of thermodynamic in these models. We ended up showing how in general only the dominant energy condition is valid in these cases while, we did test for the weak, null and strong energy conditions too. We also showed that even the generalized second law of thermodynamics maybe violated in quite a few scenarios for these primitive cutoffs. 
\\
\\
In passing we would also like to note that indeed in general we would not like such singularities or rips in cosmologies, but it is well discussed in the literature that for simple cutoffs like particle/event horizon a big rip singularity in general occurs in HDE models. Even further beyond that in recent work with the Granda-Oliveros cutoff \cite{Trivedi:2024dju}, the point of big rips being the overarching possibility for HDE models was presented very rigorously and strongly. So the question then presented itself was that if the big rip is a possibility in general, can we avoid the big rip or at least have any of its alternatives occurring consistently ? This is an interesting question and it is vital to know which cutoff would allow for big rip alternatives as well, as that would show that the particular cutoff is perhaps the best one. Again, we see that simple cutoffs like Hubble horizon,event horizon etc. do not allow for any big rip alternatives in general from various standpoints in this work. This goes to show how even though these cutoffs are a form of N-O cutoff, they are yet just very special cases while N-O cutoffs in general accommodate a lot more freedom and bigger cosmic evolution paradigms than them. Furthermore, we studied these future events in all of these cutoffs under these three HDE forms (conventional, Tsallis and Barrow) because we wanted to have a conclusive and definitive answer to which future events such cutoffs can realistically allow for or not. We wanted to test them out for these scenarios from an observational standpoint ( EOS parameters ) and also from theoretical standpoints ( validity of generalized second law, energy conditions, squared sound speed). By studying these models and seeing which future possibilities are allowed for, we wanted to show just how much more generality the NO cutoff encapsulates when compared to any other cutoff scheme. The work seems to reinforce the notion more clearly that even though all the cutoffs are just special forms of the N-O cutoff, the difference in the range of possibilities which can be accommodated in the general cutoff as compared to the other ones is indeed very big.  

\section*{Acknowledgements}
The work of S.D.O was partially supported by MICINN (Spain), project PID2019-104397GB-I00 and by the program Unidad
de Excelencia Maria de Maeztu CEX2020-001058-M, Spain. The research by M.K. was carried out at Southern Federal University with financial support from the Ministry of Science and Higher Education of the Russian Federation (State contract GZ0110/23-10-IF). The authors would also like to thank Robert Scherrer for various helpful discussions.

\printbibliography

\end{document}